# A study of M1/M2 phase cooperation in the MoV(Te,Sb)(Nb,Ta)O catalysts for propane ammoxidation to acrylonitrile

Jungwon Woo[1], Urvi Sanghavi[1], Anne Vonderheide[2], and Vadim V. Guliants[1*]

[1] Department of Biomedical, Chemical, and Environmental Engineering, University of Cincinnati, Cincinnati, OH 45221-0012

[2] Department of Chemistry, University of Cincinnati, Cincinnati, OH 45221-0037

* Corresponding author: vadim.guliants@uc.edu

**Abstract**

The bulk mixed MoV(Te,Sb)(Nb,Ta)O catalysts containing nanoscale intergrowths of so-called M1 and M2 phases display uniquely high reactivity in propane ammoxidation to acrylonitrile associated with the surface *ab* planes of the M1 phase. The current controversy surrounding this catalytic system is focused on the role of the M2 phase, which is unable to activate propane but was suggested in some studies to function in synergy with the M1 phase by efficiently converting the propylene intermediate to acrylonitrile. The present study systematically examined the catalytic behavior of pure M1 phases prepared by selective dissolution of the M2 phase in MoV(Te,Sb)(Nb,Ta)O catalysts in aqueous hydrogen peroxide. It confirmed that the MoV(Te,Sb)(Nb,Ta)O M1 phases are highly active and selective for propane and propylene ammoxidation, while the MoV(Te,Sb)(Nb,Ta)O M2 phases were active and selective in propylene ammoxidation only. Most importantly, the kinetic study of the

MoV(Te,Sb)(Nb,Ta)O M1 and M2 phases in propylene ammoxidation revealed for the very first time that that the M2 phases are significantly less active than the M1 phase in propylene ammoxidation. The findings of this study do not support the existence of the synergy effect for any M1/M2 compositional variant. Instead, the observed behavior of MoV(Te,Sb)(Nb,Ta)O catalysts was consistent with partial loss of some surface active species from the M1 phase surface during the $H_2O_2$ treatment and generation of fresh *ab* planes of the M1 phase via mechanical grinding of the $H_2O_2$-treated M1 phase. These findings provided further evidence that the M1 phase is the only phase required for the activity and selectivity of the MoV(Te,Sb)(Nb,Ta)O catalysts in propane ammoxidation to ACN.

**1. Introduction**

Replacement of olefins and aromatics feedstocks employed in selective oxidation processes by natural gas-based alkanes that are more environmentally friendly, abundant, and cheaper has received significant attention of the catalysis community in recent decades [1-6]. The ammoxidation of propane to acrylonitrile (ACN) is of particular interest as a possible alternative to the current industrial process of propylene ammoxidation providing a high-volume ACN intermediate for the manufacture of synthetic fibers, resins and rubbers [7-9]. Among different catalytic systems being investigated by academic and industrial research groups for one-step propane ammoxidation, the most promising system is the Mo-V-M (M=a combination of Nb, Te, Sb, and Ta) mixed metal oxide containing so-called M1 and M2 phases [6, 10-19].

The MoVTeNbO M1 phase was proposed to be responsible for propane activation and its selective transformation to ACN, while the MoVTeNbO M2 phase was proposed to play a secondary role by converting excess propylene intermediate that forms over the M1 surface into ACN [20, 21]. The M1 phase is capable of both propane oxidation to acrylic acid and propane ammoxidation to ACN, which were proposed to occur via the propylene intermediate [20, 22, 23]. While only the M1 phase is active and selective in propane (amm)oxidation, the presence of the M2 phase was proposed to improve the selectivity to acrylic acid [22] and the selectivity to ACN in propane (amm)oxidation [24]. The improvement in the ACN yield as a result of the MoVTeNbO M1 and M2 phase cooperation, or their synergy, was claimed in several studies [2, 19, 20, 24-26]. This synergy effect was explained by the migration of propylene intermediate formed over the surface of the M1 to the M2 phase where propylene is quickly transformed further to ACN [20]. Grasselli et al. [20] found that the synergy effect was only observed when the M1 and M2 phases, prepared separately, were ground to particle size < 5 μm and then intimately mixed. The synergy effect was also claimed for propane oxidation to acrylic acid when the M1 and M2 phases were prepared independently and then physically mixed [22]. Two possible causes of the synergy effect in the oxidation of propane to acrylic acid were proposed by Baca et al. [22]: 1) highly efficient conversion of propylene desorbed from the M1 phase into acrylic acid over the M2 phase; 2) the migration of Te from the M2 phase to the surface of the M1 phase, thus maintaining the Te concentration on the surface of the M1 which decreases during the reaction.

However, the existence of synergy effect in propane (amm)oxidation is still under debate for other chemical compositions, i.e., MoVSbNbO, MoVSbTaO, MoVSbO. For example, it was proposed that the synergy effect is not observed for propane oxidation over the MoVSb(Nb)O M1 and M2 phases [22, 27], whereas the MoVSbO M2 phase showed poor selectivity toward acrylic acid contrary to that found for the MoVTeNbO system. Furthermore, the MoVSb(Nb, Ta)O M1 and M2 phases have not been investigated for propane ammoxidation reaction.

Therefore, the first major objective of this study was to conduct a systematic investigation of catalytic reactivity of all compositional M1 and M2 phase variants as pure phases in propane and propylene ammoxidation, including exploratory synthesis and characterization of a completely new MoVSbTaO system. The second major objective of this study was to probe the existence of the synergy effect for all MoV(Te,Sb)(Nb,Ta)O M1/M2 phase mixtures in propane ammoxidation to ACN. To accomplish these objectives, we prepared well-defined M1 and M2 phases of MoVTeNbO, MoVTeTaO, MoVSbNbO, MoVSbTaO, and MoVSbO compositions employing hydrothermal (HT) and slurry evaporation (SE) synthesis methods. The pure M1, pure M2, M1/M2 physical mixtures, and as-synthesized catalysts of the MoVTeNbO, MoVTeTaO, MoVSbNbO, MoVSbTaO, and MoVSbO compositions were obtained and investigated in both propane and propylene ammoxidation to ACN.

## 2. Experimental

*2.1 Catalyst synthesis*

The MoVTeNbO catalyst with the synthesis molar ratios of the Mo:V:Te:Nb = 1:0.3:0.17:0.12 was prepared by HT method at 448 K for 48 h [23]. Ammonium molybdate (Alfa Aesar, 81-83% as $MoO_3$), vanadyl sulphate (Alfa Aesar, 99.9%), telluric acid (Alfa Aesar, 99%) and niobium (V) oxalate hexahydrate (Alfa Aesar) were used as the sources of respective elements. After hydrothermal synthesis, the catalyst precursors obtained were filtered, washed and dried at 353 K overnight. The dry precursors obtained were calcined under ultra-high purity nitrogen flow (50 ml/min) at 873 K for 2 h and ground using a mortar and pestle for 10 min to yield the as-synthesized MoVTeNbO catalyst. The details of the synthesis procedures of as-synthesized MoVTeTaO catalyst by HT method can be found in the experimental section of our earlier study [28]. The MoVTeNb(Ta)O M2 phase catalysts were prepared by slurry evaporation

(SE) method at the synthesis ratio of Mo:V:Te:Nb(Ta) = 1.00:0.31(0.3):0.27(0.37):0.08(0.06) as reported previously [25]. The selective dissolution of the MoVTe(Nb,Ta)O M2 phase was carried out by stirring the calcined as-synthesized MoVTe(Nb,Ta)O catalysts in the aqueous 30% $H_2O_2$ for 3 h at room temperature [29]. The resulting suspensions of the pure M1 phases were filtered, washed with distilled water (200 ml) and dried overnight at 353 K.

The MoVSbNbO and MoVSbTaO catalysts with synthesis molar ratios of the Mo:V:Sb:Nb(Ta)=1:0.3:0.15:0.1(0.1) were prepared by slurry evaporation as previously reported [30], but the synthesis conditions for the new MoVSbTaO system are reported in this study for the very first time. In the case of the MoVSbNb(Ta) M2 phase, the synthesis molar ratios of the Mo:V:Sb:Nb(Ta) were 1:0.33(0.3):0.3:0.1(0.1). Ammonium paramolybdate, metavanadate, and antimony trioxide were added to 45 ml of distilled water and reflexed at 363 K for 8 h. Then hydrogen peroxide (30%) was added until the black opaque suspension turned into a transparent orange solution. A second solution for the Sb/Nb and Sb/Ta systems was prepared by dissolving hydrated niobium oxide (supplied by CBMM) or tantalum (V) ethoxide, (Alfa Aesar) in aqueous solution of oxalic acid. In the case of the Sb/Ta system, hydrogen peroxide (30%) was not added. The second solution containing the niobium or tantalum source was added to the first solution mixture of the molybdenum, vanadium, and antimony sources, and the resulting solution was stirred for 10 min. The slurry was dried overnight at 483 K. The dry precursors were first calcined under air at 573 K for 4 h and then under ultra-high purity nitrogen flow (50 ml/min) at 873 K for 2 h. After calcination, the catalysts were ground using a mortar and pestle for 10 min to yield as-synthesized MoVSb(Nb,Ta)O catalysts.

The selective dissolution of the M2 phase was carried out by adding the aqueous hydrogen peroxide solution (6% $H_2O_2$) into the flask filled with the calcined catalysts and stirring for 3 h [5]. The suspension was filtered, washed with distilled water (200 ml) and dried overnight at 353K to yield the pure MoVSb(Nb,Ta)O M1 phases.

The as-synthesized MoVSbO catalyst was prepared by HT synthesis method as previously reported [27], but the synthesis conditions were slightly modified as describe below. Ammonium paramolybdate (Alfa Aesar, 81-83% as $MoO_3$), vanadyl (IV) sulfate (Alfa Aesar, 99.9%), and antimony(III) sulfate (Pro Chem, 97%,) at a Mo:V:Sb atomic ratio of 1:0.34:0.17 were stirred for 30 min in 30 ml of water. The slurry was transferred into the Teflon inner tube of a stainless steel autoclave, which was sealed and heated at 448 K for 48 h. After the hydrothermal reaction, the slurry was filtered, washed with distilled water (200 ml) and dried overnight at 353 K. The dry powdered catalyst was first calcined under air at 573 K for 20 min and then under ultra-high purity nitrogen flow (50 ml/min) at 773 K for 2 h. After calcination, the catalysts were ground using a mortar and pestle for 10 min. The dissolution of the M2 phase was carried out by adding aqueous hydrogen peroxide solution (6% $H_2O_2$) to the calcined catalysts and stirring for 3 h [29]. The suspension was filtered, washed with distilled water (200 ml) and dried overnight at 353K to yield the pure MoVSbO M1 phase.

The MoVSbO M2 phase catalyst was prepared by the slurry evaporation (SE) method at the synthesis ratio of Mo:V:Sb = 1:0.5:0.5 as previously reported [27]. The slurry was dried overnight at 483K. The dry powdered MoVSbO M2 phase catalyst was calcined under ultra-high purity nitrogen flow (50 ml/min) at 773 for 2 h before the reaction tests.

Four kinds of catalysts were prepared to probe the nature of the synergy effect in propane ammoxidation: pure M1, pure M2, physical M1/M2 mixtures, and as-synthesized catalysts containing M1/M2 intergrowths. The M1 phase was obtained after the hydrogen peroxide treatment of the calcined as-synthesized catalyst. Physical mixtures of the M1 and M2 phases were obtained by mechanically mixing the respective M1 and M2 phases in a 1:1 mass ratio and grinding them with a mortar and pestle for 10 min. The physical M1 and M2 phase mixtures were denoted as the M1/M2.

*2.2 Powder XRD*

Powder X-ray diffraction was recorded by a Siemens D500 diffractometer with Cu Ka radiation (tube voltage: 45kV, tube current: 40 mA)

*2.3 Scanning electron microscopy (SEM) and energy dispersive spectroscopy (EDS) analysis*

SEM and EDS analysis were conducted at the Advanced Materials Characterization Center at the University of Cincinnati. SEM (FEI/Philips XL 30 FEG ESEM) has a resolution of 3.5 nm at 30 kV and is equipped with the Energy Dispersive X-ray Analyzer from EDAX.

*2.4 ICP-MS elemental analysis*

Elemental analysis was performed on an Agilent 7700 Inductively Coupled Plasma Mass Spectrometer (Agilent Technologies, Santa Clara, CA). Calibration curves were examined both with and without the use of collision cell (pressurized with He, 3 ml min$^{-1}$). However, no significant differences were noted and all data reported was acquired in no gas mode. Major isotopes of all analytes of interest were measured; quantitation was performed on the highest that contained no isobaric overlap with neighboring elements. Samples were introduced with Agilent integrated auto sampler. Specific instrument parameters are given in Table 1.

**Table 1.** ICP-MS operating conditions.

| | |
|---|---|
| Plasma gas | Argon |
| Carrier gas | Argon |
| Carrier gas flow rate | 1.05 L min$^{-1}$ |
| RF Power | 1550 W |
| RF matching | 1.80 V |
| Sample depth | 10 mm |
| Nebulizer | MicroMist |
| Nebulizer Pump | 0.10 rps |
| Spray chamber temperature | 275 K |
| Make up gas flow rate | 0.12 L min$^{-1}$ |
| Total Acquisition Time | 67.584 sec |
| Replicates | 3 |
| Sweeps/Replicates | 10 |
| Integrated Time/Mass | 0.09 sec |

*2.5 Reagents and standards*

Calibration standards were purchased as 10 mg/L mixtures of Nb, Mo and Ta amongst other elements from Agilent Technologies (Agilent Technologies, Santa Clara, CA) and Spex CertiPrep (Spex CertiPrep, Metuchen, NJ). 1000 mg/L Specpure® standards for V and Te were purchased from Alfa Aesar (Alfa Aesar, Ward Hill, MA). Calibration standards were prepared by appropriate dilution of these standards with trace metal grade nitric acid (Fisher Scientific, Waltham, MA). Hydrogen peroxide, 30% was also used (Fisher Scientific, Waltham, MA). The MoVTe(Nb,Ta)O M1 phases (20 mg) were digested for 3 days at room temperature in a mixture of 49% HF (0.9 g) and 69.4% HNO$_3$ (2.5 g) and then diluted as needed for the ICP-MS analysis.

*2.6 Propane/propylene ammoxidation reaction*

The catalytic behavior of the MoV(Te,Sb)(Nb,Ta)O M1, M2, M1/M2, and as-synthesized catalysts prepared by SE and HT synthesis methods was investigated in both propane and propylene ammoxidation using a fixed bed micro-reactor equipped with an on-line GC under steady-state conditions at atmospheric pressure and 613–733 K. Powdered catalysts after calcination were ground with the mortar and pestle for 10 min and diluted with quartz sand prior to the reaction tests. The diluted catalysts were introduced into the micro-reactor, heated to the desired temperature under He flow and exposed to the reaction feed. The feed was composed of $C_3H_8$ ($C_3H_6$)/$NH_3$/$O_2$/He in the molar ratio of 5.7(5.7):8.7:17.1:68.4 at the total flow rate of 26.3 mL/min. The reactants and products were analyzed by an on-line GC system (Shimadzu 14A) equipped with a flame ionization detector and a thermal conductivity detector. The catalytic testing of the MoVTeTa(Nb) M1 phase catalysts was conducted for 48 ~ 72 h on stream for each catalyst during which these catalysts were structurally and thermally stable [31, 32]. The total carbon balances agreed within ±2%.

## 3. Results and Discussion

*3.1 Bulk characteristics of MoV(Te,Sb)(Nb,Ta)O M1, M2, M1/M2, and as-synthesized catalysts*

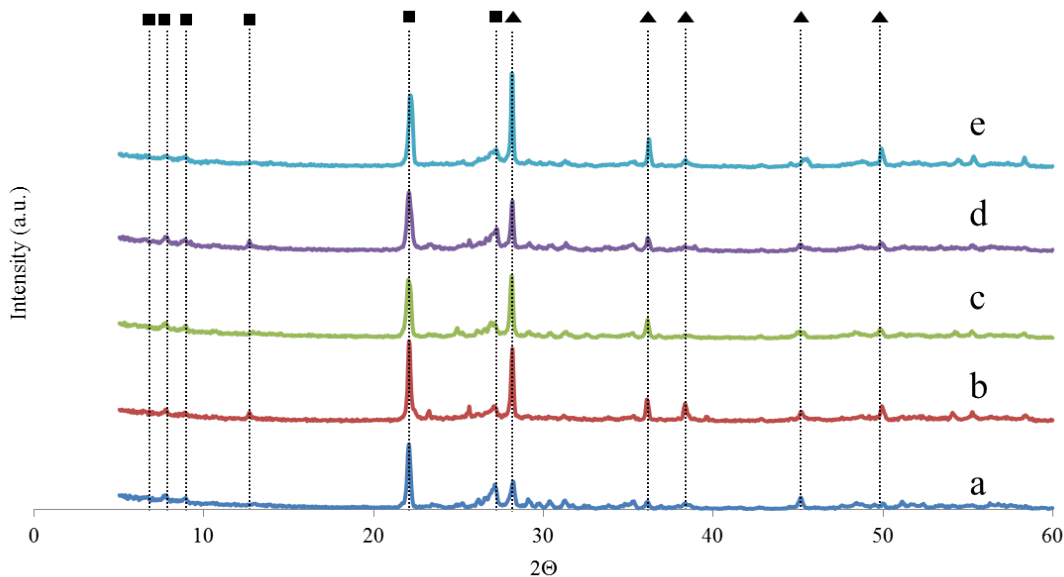

**Figure 1.** XRD patterns of the as-synthesized catalysts: a) MoVTeNbO HT; b) MoVTeTaO HT; c) MoVSbNbO SE; d) MoVSbTaO SE; e) MoVSbO HT. Symbols: (■) M1 phase; (▲) M2 phase.

The MoV(Te,Sb)(Nb,Ta)O M1, M2, and as-synthesized catalysts were characterized with XRD. Figure 1 shows XRD pattern of the MoV(Te,Sb)(Nb,Ta)O as-synthesized catalysts made by HT and SE synthesis methods. The as-synthesized MoV(Te,Sb)(Nb,Ta)O catalysts contained both the M1 and M2 phases, and their ratios varied depending on the chemical composition and synthesis method. The M1/M2 ratios in the as-synthesized MoV(Te,Sb)(Nb,Ta)O catalysts were estimated based on the analysis of their X-ray diffraction patterns [22]. For the estimation of the M1/M2 ratios of the as-synthesized MoV(Te,Sb)(Nb,Ta)O catalysts, the pure MoVTeNbO M1 and M2 phases, and three physical phase mixtures containing different mass fractions of the M1 and M2 phases (M1:M2 of 1:3, 1:1, and 3:1) were characterized by XRD. The characteristic peak areas at 2Θ = 27.1° for the M1 phase and 2Θ = 36.1° for the M2 phase of all as-synthesized catalysts were determined by the PANalytical X'Pert HighScore software. The calibration curve thus obtained was employed to estimate the M1 and M2 phase content in all as-synthesized catalysts shown in Table 2.

**Table 2.** Elemental compositions, BET surface areas, and phase ratios of the MoV(Te,Sb)(Nb,Ta)O catalysts.

| Catalysts | Preparative[a] Mo/V/Te/Ta | SEM-EDX[b] Mo/V/Te/Ta | Surface area (m$^2$/g)[c] Before H$_2$O$_2$ treatment | Surface area (m$^2$/g)[c] After H$_2$O$_2$ treatment | *M1:M2 |
|---|---|---|---|---|---|
| MoVTeNbO M1 HT | 1.00/0.3/0.17/0.12 | 1/0.29/0.12/0.20 | 5.7 | 15.3 | |
| MoVTeTaO M1 HT | 1.00/0.31/0.22/0.12 | 1/0.32/0.17/0.33 | 4.3 | 14.1 | |
| MoVSbNbO M1 SE | 1.00/0.3/0.15/0.1 | 1/0.18/0.08/0.17 | 8 | 11.4 | |
| MoVSbTaO M1 SE | 1.00/0.3/0.15/0.1 | 1/0.34/0.12/0.19 | 5.7 | 8.5 | |
| MoVSbO M1 HT | 1.00/0.34/0.16 | 1.00/0.37/0.18 | 14 | 32.8 | |
| MoVTeNbO M2 SE | 1.00/0.31/0.27/0.08 | 1/0.45/0.31/0.17 | 1.7 | | |
| MoVTeTaO M2 SE | 1.00/0.3/0.37/0.06 | 1/0.4/0.35/0.28 | 2.1 | | |
| MoVSbNbO M2 SE | 1.00/0.33/0.30/0.1 | 1/0.39/0.33/0.32 | 3.7 | | |
| MoVSbTaO M2 SE | 1.00/0.3/0.3/0.1 | 1/0.32/0.33/0.29 | 4.7 | | |
| MoVSbO M2 SE | 1.00/0.5/0.5 | 1.00/0.40/0.44 | 1.7 | | |
| MoVTeNbO M1/M2 | | | 10.2 | | |
| MoVTeTaO M1/M2 | | | 17.9 | | |
| MoVSbNbO M1/M2 | | | 11 | | |
| MoVSbTaO M1/M2 | | | 11.8 | | |
| MoVSbO M1/M2 | | | 16.2 | | |
| MoVTeNbO as-syn | 1.00/0.3/0.17/0.12 | | 11.2 | | 0.75:0.25 |
| MoVTeTaO as-syn | 1.00/0.31/0.22/0.12 | | 15.1 | | 0.48:0.52 |
| MoVSbNbO as-syn | 1.00/0.3/0.15/0.1 | | 6.7 | | 0.50:0.50 |
| MoVSbTaO as-syn | 1.00/0.3/0.15/0.1 | | 4.6 | | 0.71:0.29 |
| MoVSbO as-syn | 1.00/0.34/0.16 | | 14 | | 0.43:0.57 |

[a] Preparative composition in the slurry; [b] determined by EDS; [c] measured by BET method; HT (Hydrothermal); SE (Slurry evaporation); as-syn (as-synthesized).
*M1:M2 mass ratios were estimated as described above.

The XRD patterns of the MoV(Te,Sb)(Nb,Ta)O M1 phases, prepared by HT and SE methods, are shown in Figure 2. All MoV(Te,Sb)(Nb,Ta)O M1 phase catalysts were treated with hydrogen peroxide that selectively removed the M2 phase [29] as described in the experimental section. All of the M1 phase catalysts in Figure 2 shows a similar diffraction pattern exhibiting the characteristic peaks of the M1 phase at 2Θ=6.7, 7.8, 8.9, 22.1, 27.2, and 45° (PDF 01-073-7574) without any impurities. Figure 5.3 shows the XRD patterns of the MoV(Te,Sb)(Nb,Ta)O M2 phases. The XRD peaks at 2Θ=22, 28, 36, 45 and 50° correspond to those of the pure M2 phase (PDF 01-073-7575).

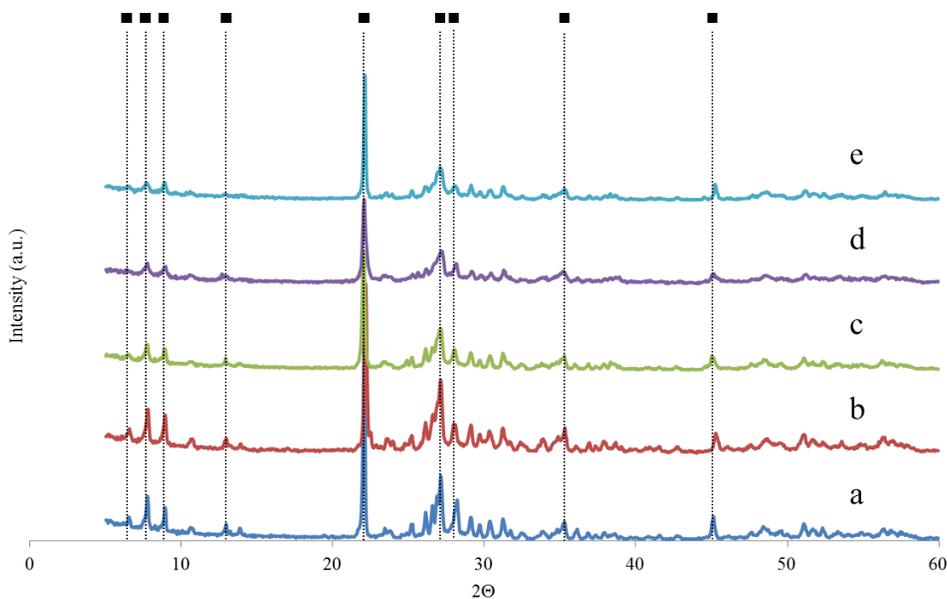

**Figure 2.** XRD patterns of the M1 phase catalysts: a) MoVTeNbO M1 HT; b) MoVTeTaO M1 HT; c) MoVSbNbO M1 SE; d) MoVSbTaO M1 SE; e) MoVSbO M1 HT. Symbol: (■) M1 phase.

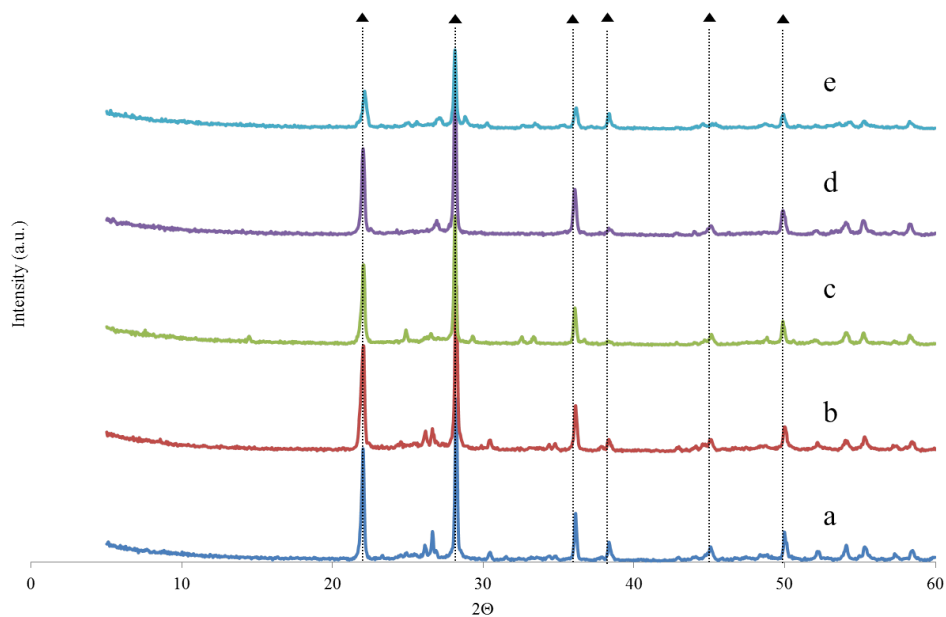

**Figure 3.** XRD patterns of the M2 phase catalysts: a) MoVTeNbO M2 SE; b) MoVTeTaO M2 SE; c) MoVSbNbO M2 SE; d) MoVSbTaO M2 SE; e) MoVSbO M2 SE. Symbol: (▲) M2 phase.

The morphology, crystal shape, and size of the MoV(Te,Sb)(Nb,Ta)O M1 and M2 phases were characterized with SEM. The SEM images of the MoV(Te,Sb)(Nb,Ta)O M1 phase catalysts in Figure 4 show the rod-like morphology regardless of synthesis methods employed and chemical compositions. It is important to note that the crystal morphology of the MoVSbNbO and MoVSbTaO M1 phases prepared by the SE method does not appear to be as ordered as that of the MoVTeNbO, MoVTeTaO, and MoVSbO M1 phases prepared by HT synthesis. This observation is consistent with the results of the previous study reported in our earlier study [28] where the MoVTeTaO M1 catalysts made by the HT method showed highly ordered crystal morphology as compared to that made by the SE method. The crystal morphology of the M2 phases is different from that of the M1 phases (Figure 5). This result is in good agreement with the results of a previous study [25] where the M2 phase particles were found to have platelet morphology. It appeared that the crystal morphology of the MoVSbO M2 phase is slightly different from that of other M2 phases (Figure 5).

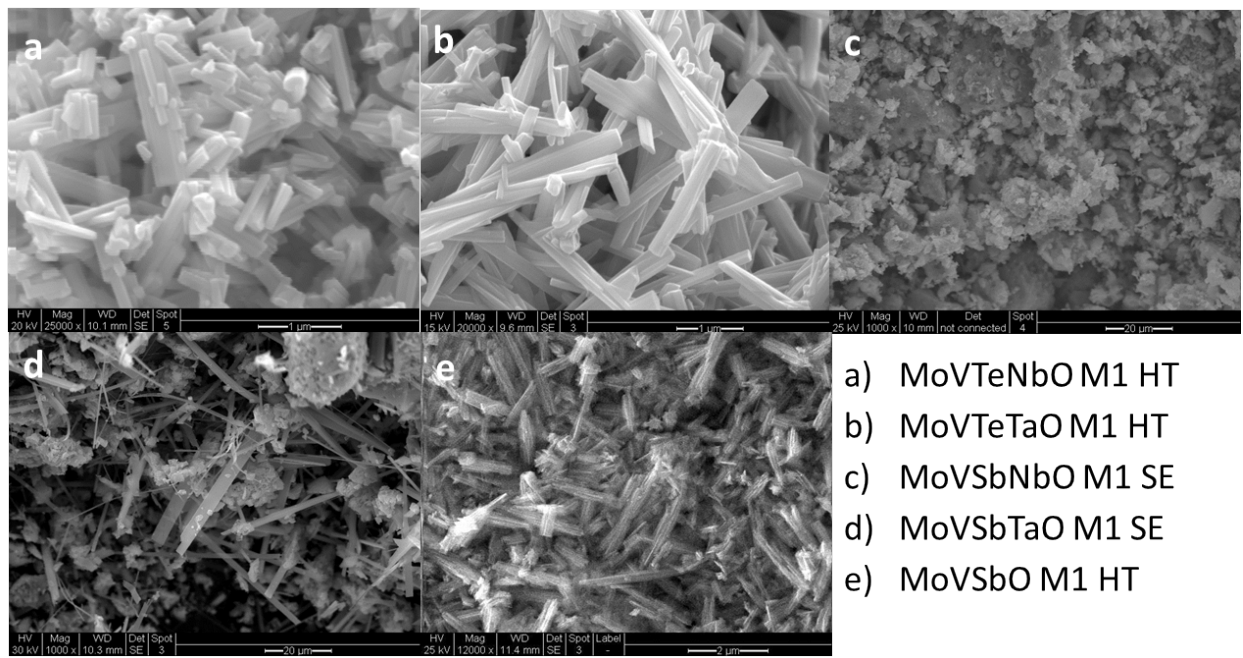

a) MoVTeNbO M1 HT
b) MoVTeTaO M1 HT
c) MoVSbNbO M1 SE
d) MoVSbTaO M1 SE
e) MoVSbO M1 HT

**Figure 4.** SEM images of the MoV(Te,Sb)(Nb,Ta)O M1 phase catalysts.

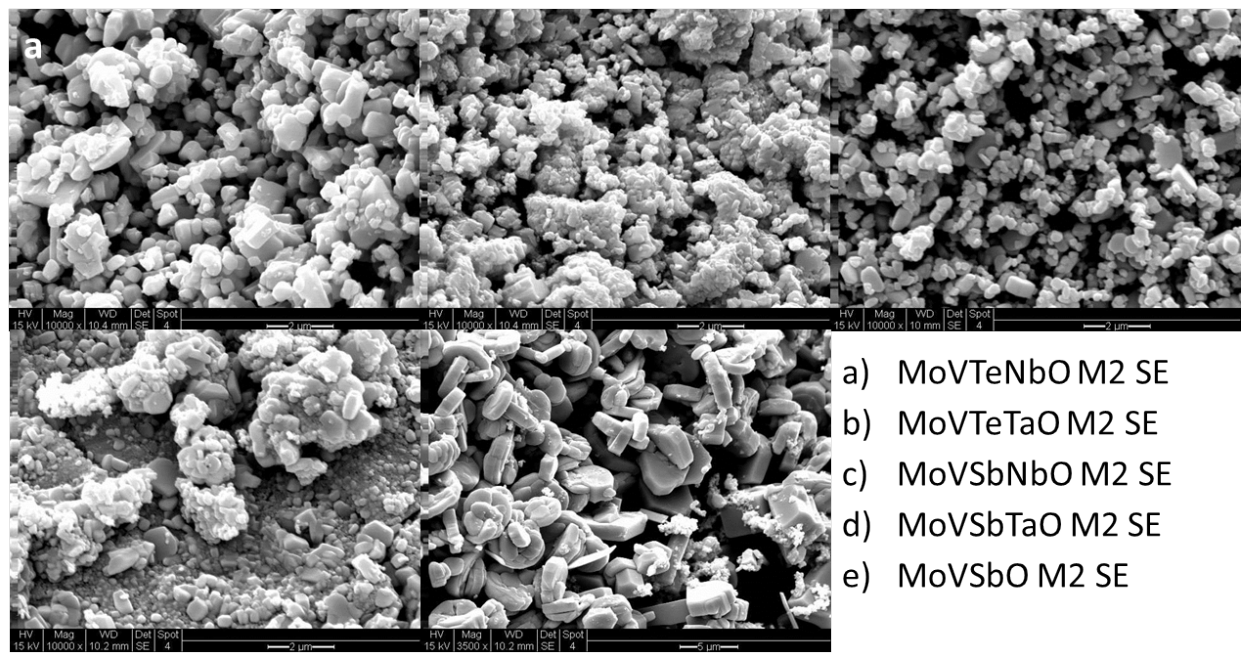

a) MoVTeNbO M2 SE
b) MoVTeTaO M2 SE
c) MoVSbNbO M2 SE
d) MoVSbTaO M2 SE
e) MoVSbO M2 SE

**Figure 5.** SEM images of the MoV(Te,Sb)(Nb,Ta)O M2 phase catalysts.

*3.2 Catalytic behavior of MoV(Te,Sb)(Nb,Ta)O M1 and M2 phase catalysts in propane and propylene ammoxidation*

*3.2.1 MoV(Te,Sb)(Nb,Ta)O M1 phase catalysts*

Five M1 phase catalysts possessing the MoVTeNbO, MoVTeTaO, MoVSbNbO, MoVSbTaO, and MoVSbO compositions were tested in propane ammoxidation. The ACN selectivity of all MoV(Te,Sb)(Nb,Ta) M1 phase catalysts was plotted against propane conversion as a function of reaction temperature in Figure 6. As expected, the three-component MoVSbO M1 phase showed the lowest selectivity to ACN as compared to other four-component catalysts, i.e., MoVTeNbO, MoVTeTaO, MoVSbNbO, MoVSbTaO, thus confirming that the fourth component, Nb or Ta, enhances the selectivity to ACN in propane ammoxidation as reported previously [23, 25, 33]. Interestingly, the Ta-containing systems, the MoVTeTaO and newly reported MoVSbTaO M1 phases, showed even higher selectivities to ACN than the Nb-containing systems, the MoVTeNbO and MoVSbNbO M1 phases, at propane conversion < 20%. However, the MoVTeNbO M1 phase showed the superior selectivity to ACN for propane conversion > 30% (Figure 6). This finding is in a good agreement with the results of a previous study [23], where the MoVTeNbO M1 phase showed the highest ACN yield among the MoVO, MoVTeO, MoVSbO, MoVSbNbO M1 phase catalysts in propane ammoxidation reaction.

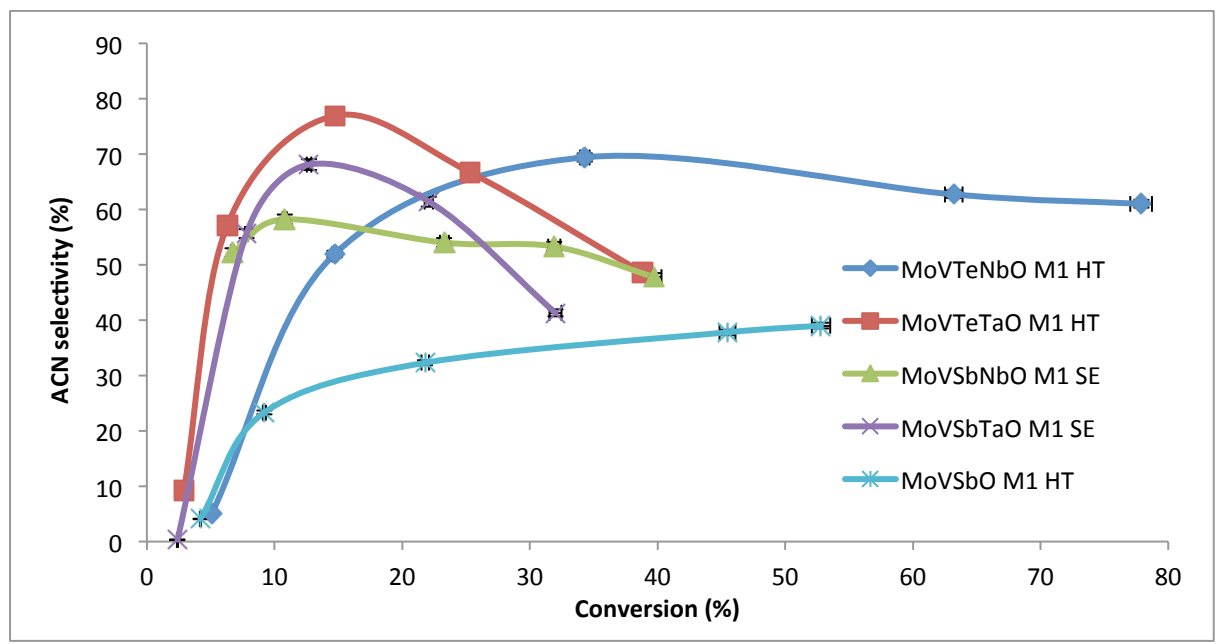

**Figure 6.** Selectivity to ACN as a function of propane conversion over the MoV(Te,Sb)(Nb,Ta)O M1 phase catalysts in propane ammoxidation; Reaction conditions: $C_3H_8:NH_3:O_2:He=5.7:8.6:17.1:68.6$ (%); total flow rate, 26.3 mL•min$^{-1}$; 0.2 g catalyst; reaction temperature: 623-733 K.

The kinetic studies of the MoV(Te,Sb)(Nb,Ta)O M1 phase catalysts were conducted in order to further understand their catalytic behavior in propane ammoxidation to ACN. According to previous kinetic studies of the MoVTeNbO M1 phase, the propane consumption over the MoVTeNbO M1 phase under propane ammoxidation conditions was found to be a *first-order irreversible* reaction [23, 34]. The kinetic studies of the MoVTeTaO M1 phase in propane ammoxidation reported in our earlier study [28] supported this conclusion and suggested that all MoV(Te,Sb)(Nb,Ta)O M1 phase catalysts in propane ammoxidation would also follow the *first-order irreversible* reaction kinetics. The linear nature of Arrhenius plots of the first-order irreversible reaction rate constant, k''($C_3H_8$), for all MoV(Te,Sb)(Nb,Ta)O M1 phase catalysts shown in Figure 1 of Supporting Information confirmed that propane consumption over all MoV(Te,Sb)(Nb,Ta)O M1 phase catalysts during propane ammoxidation is the 1$^{st}$ order reaction.

Figure 1 of Supporting Information further showed that the MoVTeNbO M1 phase is the most active catalyst, while the activity in propane ammoxidation reaction decreased in the following order: MoVTeNbO M1 > MoVSbO M1 > MoVSbNbO M1 > MoVTeTaO M1 > MoVSbTaO M1. The MoVTeTaO and MoVSbTaO M1 phases showed relatively higher selectivity to ACN than the MoVTeNbO and MoVSbNbO M1 phases (Figure 6), while the MoV(Te,Sb)TaO M1 phases were less active than the MoV(Te,Sb)Nb M1 phases (Figure 1 of Supporting Information). These findings clearly suggested that the MoVTeNbO M1 phase is the overall best catalyst in propane ammoxidation to ACN based on its relatively high selectivity to ACN and highest activity among all M1 variants investigated in this study.

*3.2.2 Propylene ammoxidation over MoV(Te,Sb)(Nb,Ta)O M1 and M2 phase catalysts*

The both MoVTeNbO M1 and M2 phases were previously shown to be active for propylene (amm)oxidation [1, 19, 20, 22]. However, few studies directly compared the catalytic performance of the M1 and M2 phases in propylene ammoxidation [20, 33], whereas other chemical compositions, e.g, MoVTeTaO, MoVSbNbO, MoVSbTaO, MoVSbO, have not been investigated. Therefore, all MoV(Te,Sb)(Nb,Ta)O M1 and M2 phase catalysts were systematically investigated in propylene ammoxidation in this study.

The ACN selectivity of all MoV(Te,Sb)(Nb,Ta)O M1 and M2 phase catalysts was plotted against propylene conversion as a function of reaction temperatures in Figure 7 and 8, respectively. All MoV(Te,Sb)(Nb,Ta) M1 phase catalysts were found to be selective to ACN, depending on the chemical composition. These findings are in good agreement with the results of previous studies [1, 2, 35], where the M1 phase contained the active surface sites capable of both propane and propylene activation to ACN. Interestingly, the MoVSbNbO and MoVSbTaO M1 phases displayed ~ 80 mol. % selectivity to ACN, whereas the MoVTeNbO and MoVTeTaO M1 phases shows 60 ~ 65 mol. % selectivity to ACN at ~15 % propylene conversion (Figure 7). This observation suggested that the Sb and Te cations, incorporated into the M1 lattice behave differently during propylene ammoxidation, although both of them were proposed to have the same function, i.e., activating propylene through α-hydrogen abstraction to ACN.

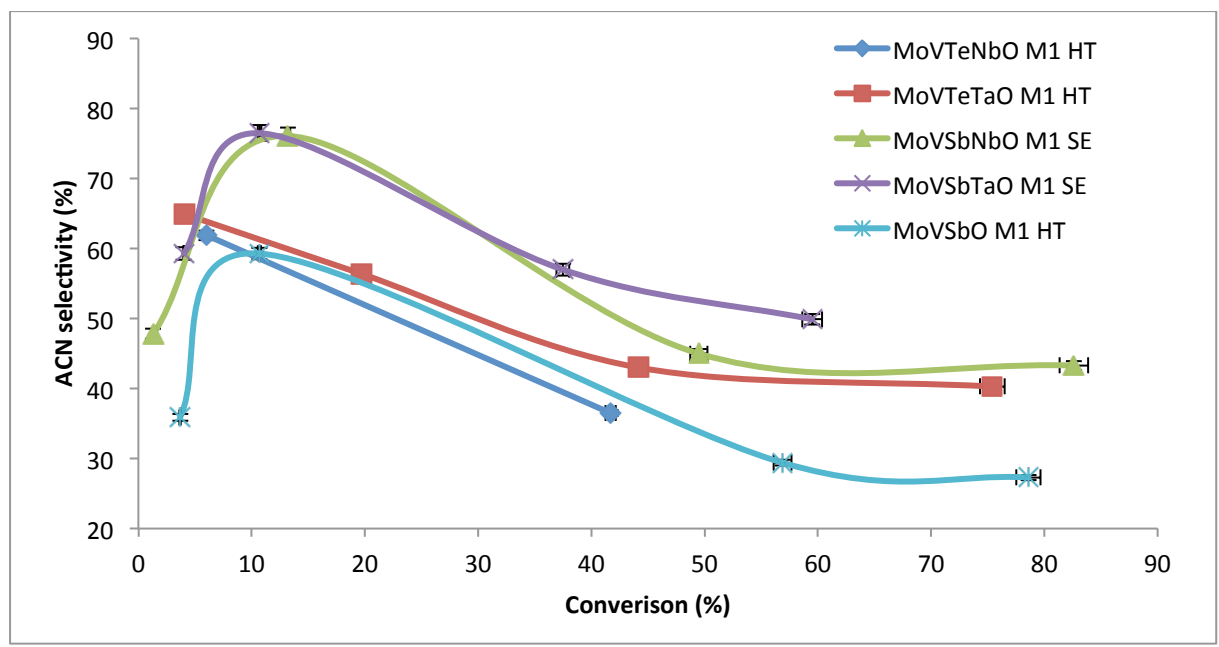

**Figure 7.** Selectivity to ACN as a function of propylene conversion over the MoV(Te,Sb)(Nb,Ta)O M1 phase catalysts in propylene ammoxidation; Reaction conditions: $C_3H_6:NH_3:O_2:He=5.7:8.6:17.1:68.6$ (%); total flow rate, 26.3 mL•min$^{-1}$; 0.2 g catalyst; reaction temperature: 613-673 K.

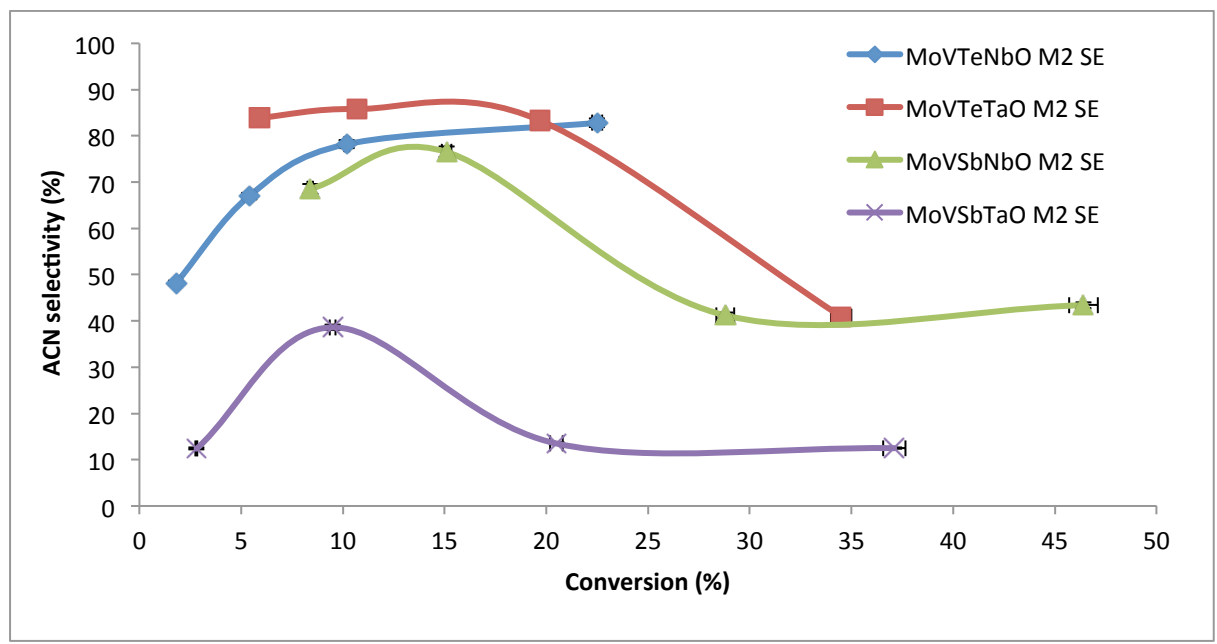

**Figure 8.** Selectivity to ACN as a function of propylene conversion over the MoV(Te,Sb)(Nb,Ta)O M2 phase catalysts in propylene ammoxidation; Reaction conditions: $C_3H_6$:$NH_3$:$O_2$:He=5.7:8.6:17.1:68.6 (%); total flow rate, 26.3 mL•min$^{-1}$; 0.2 g catalyst; reaction temperature: 653-713 K.

All MoV(Te,Sb)(Nb,Ta)O M2 phase catalysts were found to possess different selectivities as shown in Figure 8, with the exception of the MoVSbO M2 phase which was found to be inactive in propylene ammoxidation under the experimental conditions of this study. The MoVTeNbO, MoVTeTaO, MoVSbNbO M2 phases displayed relatively high selectivity toward ACN (70 ~ 80 mol. %) while the MoVSbTaO shows relatively low selectivity to ACN (~ 40 mol. %) in propylene ammoxidation at low propylene conversion (~ 15 %).

Similar to the catalytic activity of the MoV(Te,Sb)(Nb,Ta)O M1 phases in propane ammoxidation, the propylene consumption over the MoV(Te,Sb)(Nb,Ta)O M1 and M2 phase catalysts in propylene ammoxidation was assumed to be the *first-order irreversible* reaction [36]. The Arrhenius plots of the first-order irreversible reaction rate constant, k"($C_3H_6$), for the M1 and M2 of the MoV(Te,Sb)(Nb,Ta)O catalysts are shown in Figures 2 and 3 of Supporting Information, respectively. The linear nature of the Arrhenius plots of the MoV(Te,Sb)(Nb,Ta)O M1 and M2 phase catalysts reported in Figures 2 and 3 of Supporting Information further supported our assumption that the propylene consumption over the MoV(Te,Sb)(Nb,Ta)O M1 and M2 phase catalysts in propylene ammoxidation is indeed a 1$^{st}$ order reaction.

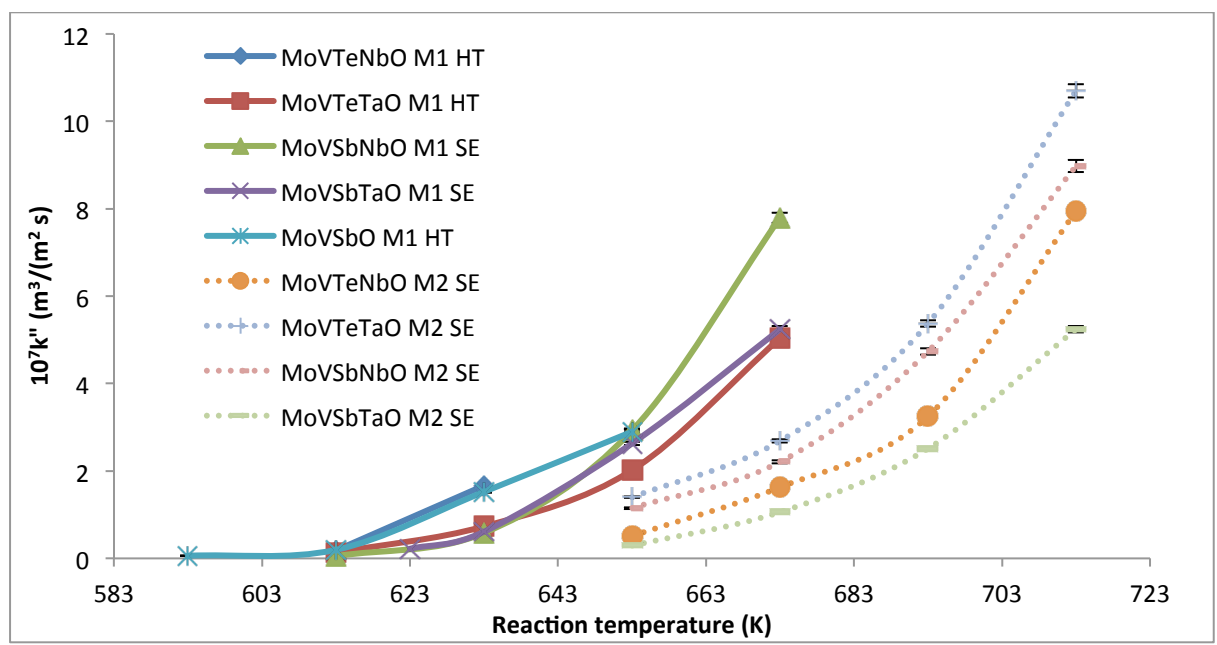

**Figure 9.** Reaction rate constants of propylene consumption, k"($C_3H_6$), vs. reaction temperature for the MoV(Te,Sb)(Nb,Ta)O M1 and M2 phase catalysts in propylene ammoxidation; Reaction conditions: $C_3H_6$:$NH_3$:$O_2$:He=5.7:8.6:17.1:68.6; total flow rate, 26.3 mL•min$^{-1}$; 0.2 g catalyst.

Figure 9 shows a comparison of the reaction rate constants for propylene consumption, k"($C_3H_6$), over all MoV(Te,Sb)(Nb,Ta)O M1 and M2 phases in propylene ammoxidation as a function of reaction temperature in the range of 613-713 K. The k"($C_3H_6$) values for all MoV(Te,Sb)(Nb,Ta)O M1 phase catalysts showed some dependence on the chemical compositions, while the k"($C_3H_6$) values for all MoV(Te,Sb)(Nb,Ta)O M2 phases displayed significant dependence on the chemical compositions in propylene ammoxidation (Figure 9). The

k"($C_3H_6$) of MoV(Te,Sb)(Nb,Ta)O M2 phases decreased in the following order: MoVTeTaO > MoVSbNbO > MoVTeNbO > MoVSbTaO. It is highly important to note that k"($C_3H_6$) of MoV(Te,Sb)(Nb,Ta)O M1 phases were dramatically higher than those of respective MoV(Te,Sb)(Nb,Ta)O M2 phases at the same reaction temperature. *These findings strongly indicated that all M1 phases are much more active than all M2 phases in converting propylene to ACN regardless of their chemical composition.* In a previous study, Holmberg et al. [23] prepared the MoVTeNbO catalysts that contained a mixture of the M1, M2, and rutile phase, and correlated the catalytic activity, expressed as % conversion per $m^2$ of surface area, and the selectivity to ACN in propylene ammoxidation with the content of these phases, expressed as the ratios of major XRD reflections of each phase. They found all three phases to have similar activity, with M1 being slightly more active than the M2 and rutile phase. However, the results of our systematic study that employed pure, well-defined M1 and M2 phases provided direct evidence that all M1 phases were significantly more active than all M2 phases in propylene ammoxidation, as expressed by their $1^{st}$ order irreversible reaction rate constants normalized to their BET surface area as discussed further below for specific MoV(Te,Sb)(Nb,Ta)O compositions. Our results are further supported by the earlier findings of Ishchenko et al. [37], which showed the MoVTeNbO M1 phase to possess much higher activity as compared to the M2 phase in propylene oxidation reaction.

*3.3 Synergy of MoV(Te,Sb)(Nb,Ta)O M1 and M2 phases in propane ammoxidation*

*3.3.1 MoVTeNbO M1, M1/M2, and as-synthesized catalysts in propane ammoxidation*

The improvement of the yield of acrylic acid and ACN in propane (amm)oxidation over a mixture of the MoVTeNbO M1 and M2 phases were reported by several groups [2, 19, 22, 24, 25]. Holmberg et al. [20] suggested that the increase in the yield of ACN as a result of the

M1/M2 synergy is explained by the conversion of unreacted propylene intermediate migrating from the M1 to M2 phase [1, 2]. Baca et al. [22] also proposed the participation of the M2 phase in transformation of propylene formed on M1 phase as one of the origins of the synergy effect.

However, a recent study from our group indicated that the MoVTeNbO M1 phase is the only active and selective phase in propane ammoxidation to ACN [33]. This conclusion is based on the findings that the MoVTeNbO M1 phase was more efficient in propane ammoxidation at longer reactor residence times and higher propane conversion [33]. In addition, Ishchenko et al. [37] proposed that the M1 phase is sufficient for the oxidation of propane to acrylic acid while the M2 phase is undesirable due to its lower activity and selectivity as compared to the M1 phase. Therefore, it is important to elucidate the existence of the synergy effect between the MoVTeNbO M1 and M2 phases in propane ammoxidation.

Therefore, a systematic study of propane and propylene ammoxidation over (1) pure M1, (2) pure M2, (3) M1/M2 physical mixtures, and (4) as-synthesized MoVTeNbO catalysts was conducted in order to probe the existence of the synergy effect for this catalytic system. The M1/M2 catalyst is a physical mixture of the M1 and M2 phases (M1:M2=1:1 mass ratio) described in experimental section. The M1:M2 content (wt. %) in the as-synthesized MoVTeNbO catalyst was estimated to be 0.75:0.25 (Table 2). The selectivities to ACN for the MoVTeNbO M1, M1/M2, and as-synthesized catalysts in propane ammoxidation are plotted as a function of propane conversion in Figure 4 of Supporting Information. These results suggested improved selectivities to ACN for the M1/M2 and as-synthesized catalysts as compared to the pure M1 at low propane conversion (< 20 %) during propane ammoxidation (Figure 4 of Supporting Information).

However, the selectivities to ACN of all catalysts became similar at propane conversion > 50 %. This result is similar to that observed in the earlier study [33] where the as-synthesized MoVTeNbO catalyst (75 % M1 and 25 % M2) was found to be more efficient at low to moderate propane conversion (~ 50 %), but the M1 phase alone became more efficient as propane conversion increased above 50 % during its ammoxidation. Therefore, these results suggested that the M1 phase alone is sufficiently active in propane ammoxidation, while some improvement in the selectivity to ACN was observed for the M2-containing catalysts at low propane conversion.

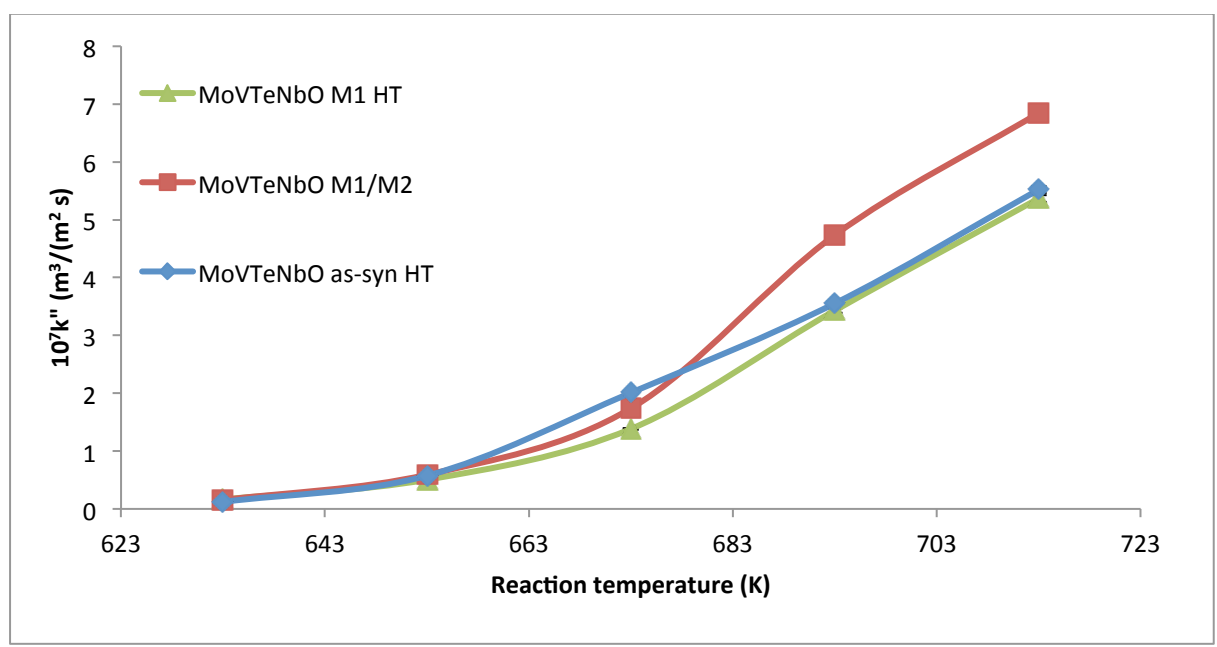

**Figure 10**. Reaction rate constants of propane consumption, k''($C_3H_8$), vs. reaction temperature for the MoVTeNbO M1, M1/M2, and as-synthesized catalysts in propane ammoxidation; Reaction conditions: $C_3H_8$:$NH_3$:$O_2$:He=5.7:8.6:17.1:68.6; total flow rate, 26.3 mL•min$^{-1}$; 0.2 g catalyst.

The kinetic studies of the MoVTeNbO M1, M1/M2, and as-synthesized catalysts were conducted to further understand the observed improvement of the ACN selectivity at low propane conversion reported in Figure 4 of Supporting Information. The propane reaction rate constants, k"($C_3H_8$), of the MoVTeNbO M1, M1/M2, and as-synthesized catalysts are plotted against the reaction temperature in Figure 10. If one assumes that the M2 phase is inactive towards propane and only converts propylene to ACN, then the pure M1 phase catalyst was expected to display the highest reaction rate constant, k"($C_3H_8$), as compared to the M1/M2 and as-synthesized catalysts, which contained a significant fraction of the inactive M2 phase. For instance, the M1/M2 catalyst contained 50 wt. % M2, while the as-synthesized catalyst contained 25 wt. % M2 (Table 2). However, k"($C_3H_8$) of the M1/M2 catalyst was higher than that of the M1 catalyst above 653 K as shown in Figure 10, which is against this expectation. This higher k"($C_3H_8$) of the M1/M2 catalyst as compared to that of the pure M1 phase suggested three possible scenarios: (1) the M2 phase participation in propane activation; (2) detrimental impact of the $H_2O_2$ treatment employed to remove the M2 phase on the surface chemistry of the resulting pure M1 phase; and (3) enhanced catalytic activity of the M1 phase due to preferential exposure of fresh *ab* planes as a result of crushing the M1/M2 phase mixture to improve its interfacial contact. The results of the previous [20-22, 38-40] and present study clearly demonstrated that the M2 phase is not capable of propane activation. Therefore, the only reasonable explanations for the high k"($C_3H_8$) value of the M1/M2 catalyst are scenarios 2 and 3 above.

As described in the experimental section, the as-synthesized M1/M2 catalyst was treated by $H_2O_2$ to selectively dissolve the M2 phase for preparation of the pure M1 phase. We speculate that the $H_2O_2$ treatment not only dissolves the bulk M2 phase but also removes some surface component from the remaining M1 phase, which has a detrimental impact on its catalytic activity. Therefore, the M1 phase obtained after the $H_2O_2$ treatment is expected to be different from the M1 phase present in the as-synthesized catalyst because it might lacks some surface metal oxide species important for its activity towards propane. The detrimental impact of the $H_2O_2$ treatment on the M1 phase is further discussed in section 3.4.

On the other hand, while the pure M1 phase after the $H_2O_2$ treatment was not crushed prior to its catalytic testing, this pure M1 was thoroughly ground together with the M2 phase in a mortar and pestle for 10 min for the preparation of the M1/M2 catalyst because the synergy effect was previously claimed only when these phases were mixed on micro-scale to improve their interfacial contact (for particles <5 μm) [20]. It is well known that this grinding preferentially exposes fresh *ab* planes of the M1, which are proposed to contain active and selective surface sites for propane ammoxidation [24, 41, 42]. A study of the MoVTeNbO M1 phase that selectively exposed *ab* planes also provided additional evidence that the *ab* planes may contain the active and selective sites for propane ammoxidation [43]. The higher BET surface area of the M1/M2 catalyst than the M1 phase reported in Table 2 further supports the proposal that the M1/M2 grinding increased the surface area of freshly exposed *ab* planes of the M1 phase. Therefore, observed high k"($C_3H_8$) of the M1/M2 catalyst for propane ammoxidation shown in Figure 10 can be explained by the additionally exposed *ab* planes exposed by grinding as reported previously which is a well-known phenomenon reported as the grinding effect [33]. This grinding effect can explain the improved selectivity to ACN over the M1/M2 catalyst at low propane conversion (Figure 4 of Supporting Information), while the removal of some surface active species from the M1 surface by $H_2O_2$ may explain lower catalytic activity of the pure M1 phase thus obtained.

*3.3.2 MoVTeNbO M1, M2, M1/M2, and as-synthesized catalysts for propylene ammoxidation*

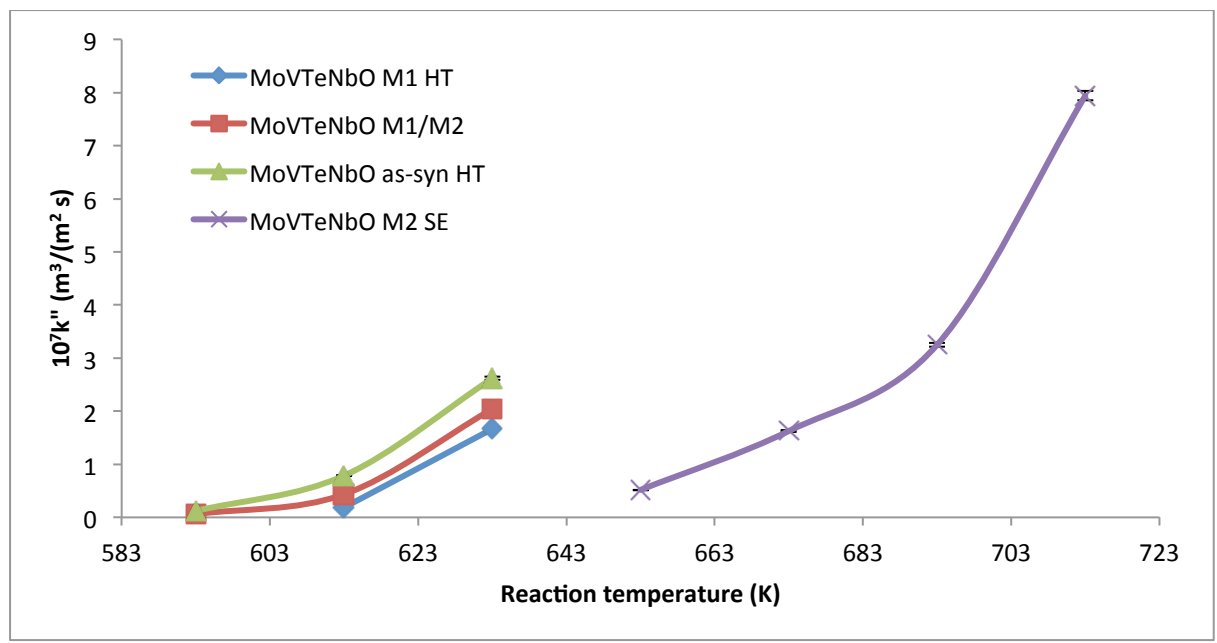

**Figure 11**. Reaction rate constants of propylene consumption, k" ($C_3H_6$), vs. reaction temperature for the MoVTeNbO M1, M2, M1/M2, and as-synthesized catalysts in propylene ammoxidation; Reaction conditions: $C_3H_6$:$NH_3$:$O_2$:He=5.7:8.6:17.1:68.6; total flow rate, 26.3 mL•min$^{-1}$; 0.2 g catalyst.

The kinetic studies of the MoVTeNbO M1, M2, M1/M2, and as-synthesized catalysts were performed in order to further understand their catalytic behavior in propylene ammoxidation. The propylene consumption rate constants, k"($C_3H_6$), of the M1, M2, M1/M2, and as-synthesized catalysts were plotted against the reaction temperature in Figure 11. It appeared that all M1-containing phases are more active toward propylene than the pure M2 even though the reaction temperature ranges do not overlap under the reaction conditions of this kinetic study (Figure 11). Furthermore, the k"($C_3H_6$) values of the M1, M2, M1/M2, and as-synthesized catalysts were estimated based on Arrhenius fits (Figure 2 and 3 of Supporting information) at the same temperature, 673 K, and these k"($C_3H_6$) values, normalized to the k"($C_3H_6$) value of the M2 phase, were compared; M1 : M1/M2 : as-syn : M2 = 65 : 44 : 31 : 1. These results clearly indicated that the MoVTeNbO M2 phase is significantly less active than the

MoVTeNbO M1 phase although the M2 phase may be somewhat more selective than the M1 phase toward ACN in propylene ammoxidation (Figures 7 and 8). Slightly lower selectivity of the pure M1 phase during propylene ammoxidation as compared to the pure M2 phase may be also explained by the partial removal of some surface active species from the M1 surface after the $H_2O_2$ treatment. Most importantly, these findings strongly suggested the absence of the M1/M2 synergy effect because the M2 phase is inactive in propane ammoxidation and significantly less active than the M1 phase in propylene ammoxidation. Moreover, the improved selectivity to ACN observed for the MoVTeNbO M1/M2 catalyst in propane ammoxidation can instead be explained by crushing the M1 phase, which selectively exposes fresh surface *ab* planes proposed to be active and selective in propane (amm)oxidation [24, 41, 44].

*3.3.3 MoVTeTaO M1, M1/M2, and as-synthesized catalysts in propane ammoxidation*

The selectivities to ACN during propane ammoxidation over the MoVTeTaO M1, M1/M2, and as-synthesized catalysts are plotted as a function of propane conversion in Figure 5 of Supporting Information. The M1/M2 and as-synthesized catalysts (Figure 5 of Supporting Information) showed slightly higher selectivity to ACN than the M1 phase.

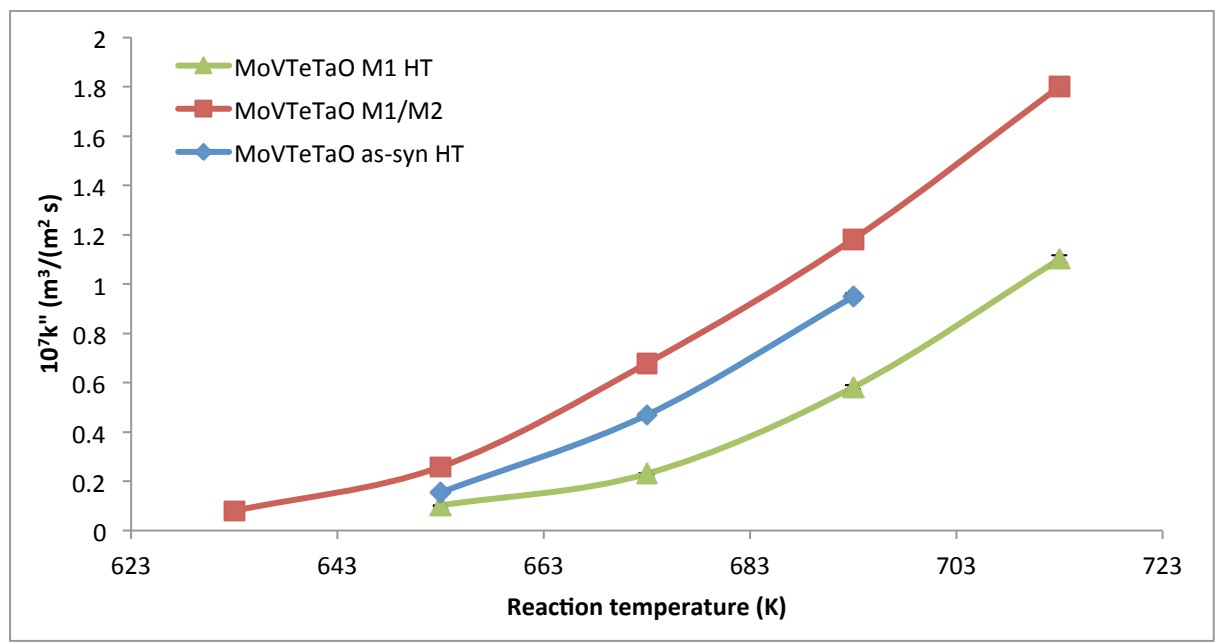

**Figure 12**. Reaction rate constants of propane consumption, k"($C_3H_8$), vs. reaction temperature for the MoVTeTaO M1, M1/M2, and as-synthesized catalysts in propane ammoxidation; Reaction conditions: $C_3H_8:NH_3:O_2:He$=5.7:8.6:17.1:68.6; total flow rate, 26.3 mL•min$^{-1}$; 0.2 g catalyst.

We further performed the kinetic studies of the MoVTeTaO M1, M1/M2, and as-synthesized catalysts in propane ammoxidation (Figure 12) in order to understand the greater selectivity to ACN observed for the MoVTeTaO M1/M2 and as-synthesized catalysts in propane ammoxidation as compared to pure M1 phase (Figure 5 of Supporting Information). The propane consumption rate constants, k"($C_3H_8$), of the MoVTeTaO M1, M1/M2, and as-synthesized catalysts are plotted against the reaction temperature in Figure 12. Similar to the MoVTeNbO M2 phase, the MoVTeTaO M2 phase was also found to be inactive for propane ammoxidation in this study, which agrees well with the results of Grasselli et al. [25]. The high k"($C_3H_8$) values of the MoVTeTaO M1/M2 catalyst for propane ammoxidation (Figure 12) can also explained by the additionally exposed *ab* planes produced by grinding as reported previously [33]. This grinding effect can explain the improved selectivity to ACN over the M1/M2 catalyst (Figure 5 of Supporting Information), while the removal of some surface active species from the M1 surface by $H_2O_2$ treatment may explain lower catalytic activity of the pure M1 phase thus obtained.

*3.3.4 MoVTeTaO M1, M2, M1/M2, and as-synthesized catalysts in propylene ammoxidation*

The kinetic studies of the MoVTeTaO M1, M2, M1/M2, as-synthesized catalysts in propylene ammoxidation were conducted in order to understand somewhat higher selectivity to ACN for the MoVTeTaO M1/M2 and as-synthesized catalysts as compared to pure M1 phase (Figure 5 of Supporting Information). The propylene consumption rate constants, k"($C_3H_6$), for the MoVTeTaO catalysts were plotted against the reaction temperature in Figure 13. Similar to that observed for the MoVTeNbO catalysts, the MoVTeTaO M1, M1/M2, and as-synthesized catalysts showed similar propylene consumption rate constants, k"($C_3H_6$), reported in Figure 13. At the same reaction temperature (e.g., 653 K), the k" ($C_3H_6$) of the M2 phase is lower than that of all M1-containing catalysts. This result indicated that the M2 phase would produce ACN at a lower yield than the M1 phase due to the low activity of the M2 phase in propylene ammoxidation, even though it is somewhat more selective to ACN (Figures 7 and 8). Therefore, these results indicated that the MoVTeTaO M1 phase is significantly more efficient than the MoVTeTaO M2 phase in terms of converting the propylene intermediate to ACN. This conclusion further suggests the absence of the synergy effect for the MoVTeTaO M1/M2 phases in propane ammoxidation to ACN.

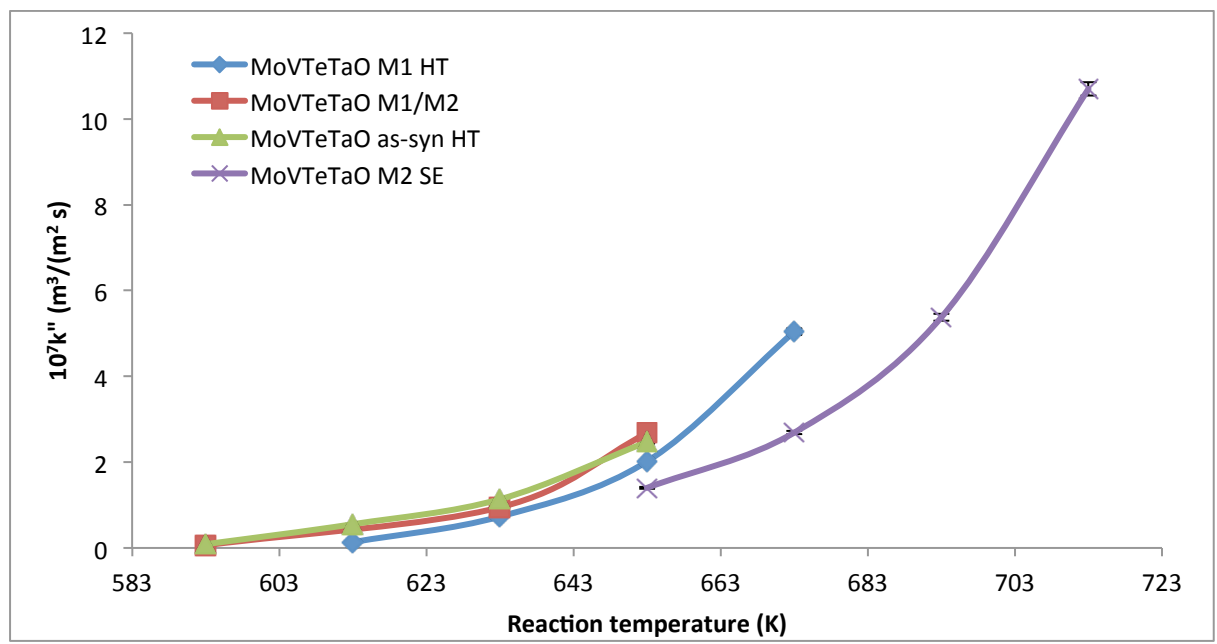

**Figure 13**. Reaction rate constants of propylene consumption, k" ($C_3H_6$), vs. reaction temperature for the MoVTeTaO M1, M2, M1/M2, and as-synthesized catalysts in propylene ammoxidation; Reaction conditions: $C_3H_6$:$NH_3$:$O_2$:He=5.7:8.6:17.1:68.6; total flow rate, 26.3 mL•min$^{-1}$; 0.2 g catalyst.

*3.3.5 MoVSbNbO M1, M1/M2, and as-synthesized catalysts in propane ammoxidation*

Baca et al. [22] studied the catalytic behavior of the MoVTe(Sb)NbO M1 and M2 phases in the oxidation of propane to acrylic acid and found no synergy effect between the M1 and M2 phase. However, the synergy effect for propane ammoxidation over the MoVSbNbO catalysts has not been investigated. Therefore, the MoVSbNbO M1, M1/M2, and as-synthesized catalysts were investigated in order to probe the existence of the synergy effect for the MoVSbNbO system in propane ammoxidation to ACN. The selectivities to ACN in propane ammoxidation for all MoVSbNbO catalysts were plotted as a function of propane conversion in Figure 6 of Supporting Information, which demonstrated enhanced selectivity to ACN for the M1/M2 and as-synthesized catalysts as compared to the pure M1 phase obtained after the $H_2O_2$ treatment without subsequent grinding.

We further performed the kinetic studies of the MoVSbNbO M1, M1/M2, and as-synthesized catalysts in propane ammoxidation in order to understand the improved ACN selectivity for the M1/M2 and as-synthesized catalysts in propane ammoxidation as compared to the pure M1 phase (Figure 6 of Supporting Information). The $k"(C_3H_8)$ values of the M1, M1/M2, and as-synthesized catalysts are plotted against the reaction temperature in Figure 14. Similar to the MoVTe(Nb,Ta)O M2 phase, the MoVSbNbO M2 phase was also found to be inactive for propane ammoxidation in this study. Since M2 phase is inactive towards propane, the pure M1 phase catalyst was presumed to display the highest $k"(C_3H_8)$ value as compared to the M1/M2 (50% M1 and 50% M2) and as-synthesized catalysts (50% M1 and 50% M2) according to Table 2. However, the M1/M2 and as-synthesized catalyst showed higher $k"(C_3H_8)$ values than the pure M1 phase above 653 K as shown in Figure 14. As explained in section 3.3.1, three possible scenarios are available for explaining this high $k"(C_3H_8)$ of the M1/M2 catalyst and low $k"(C_3H_8)$ of the pure M1 catalyst. The first scenario is excluded because the MoVSbNbO M2 was found to be incapable of propane activation in present study. Therefore, the only reasonable explanations for the high $k"(C_3H_8)$ values of the M1/M2 catalyst are scenarios 2 and 3 as described in section 3.3.1. Therefore, the lack of some surface metal oxide species important for its activity towards propane during the $H_2O_2$ treatment according to scenario 2 can explain the low activity of the pure M1 phase as compared to the M1/M2 and as-synthesized catalysts in propane ammoxidation to ACN (Figure 14), while the observed high $k"(C_3H_8)$ of the M1/M2 catalyst for propane ammoxidation shown in Figure 14 can be explained by the grinding effect based on scenario 3.

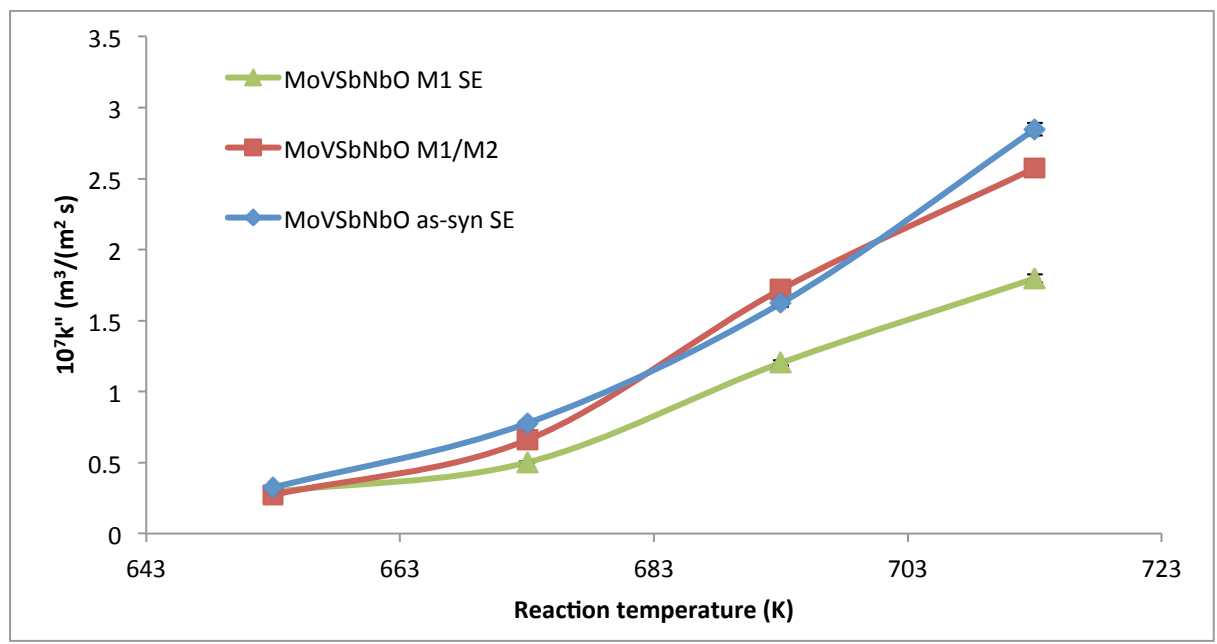

**Figure 14.** Reaction rate constants of propane consumption, k"($C_3H_8$), vs. reaction temperature for the MoVSbNbO M1, M1/M2, and as-synthesized catalysts in propane ammoxidation; Reaction conditions: $C_3H_8$:$NH_3$:$O_2$:He=5.7:8.6:17.1:68.6; total flow rate, 26.3 mL•min$^{-1}$; 0.2 g catalyst.

*3.3.6 MoVSbNbO M1, M2, M1/M2, and as-synthesized catalysts in propylene ammoxidation*

The MoVSbNbO M1, M2, M1/M2, and as-synthesized catalysts were tested in order to provide additional insights into catalytic behavior of the M1 and M2 phases in propylene ammoxidation. It was found that both the MoVSbNbO M1 and M2 phases were active and selective to ACN in propylene ammoxidation similar to that observed for the MoVTe(Nb,Ta)O system. Moreover, the MoVSbNbO M1 and M2 phases showed very similar ACN selectivity behavior as a function of propane conversion (Figure 7 and 8).

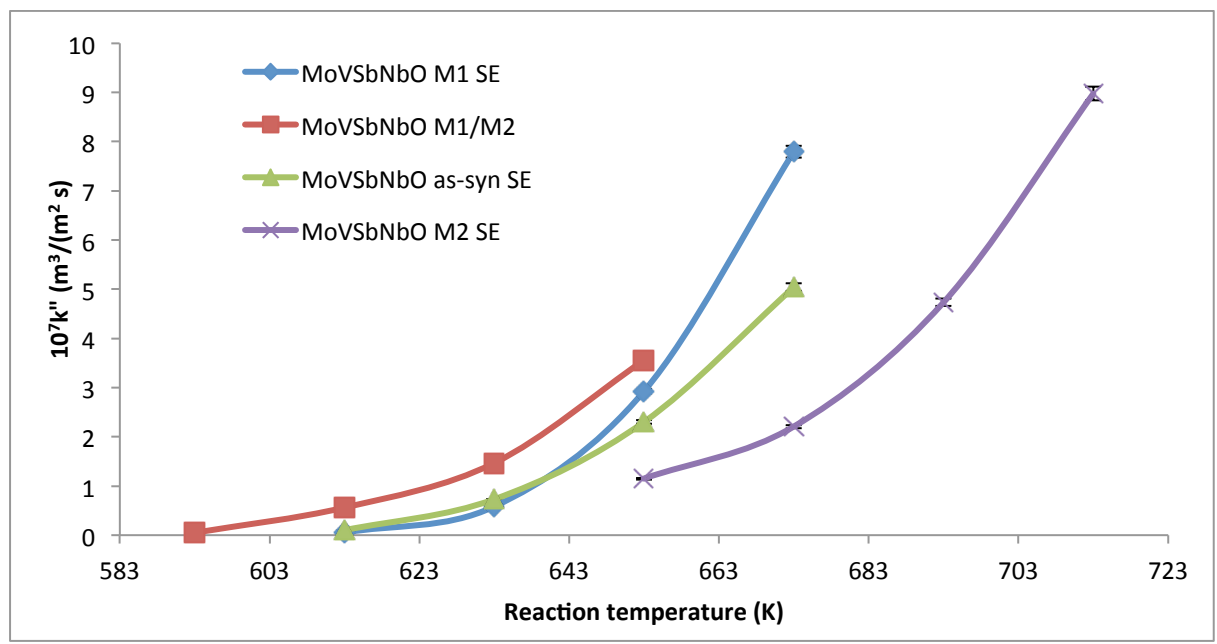

**Figure 15**. Reaction rate constants of propylene consumption, k" ($C_3H_6$), vs. reaction temperature for the MoVSbNbO M1, M2, M1/M2, and as-synthesized catalysts in propylene ammoxidation; Reaction conditions: $C_3H_6$:$NH_3$:$O_2$:He=5.7:8.6:17.1:68.6; total flow rate, 26.3 mL•min$^{-1}$; 0.2 g catalyst.

The k"($C_3H_6$) values for the MoVSbNbO catalysts were plotted against the reaction temperature in Figure 15. Figure 15 indicated that the k" ($C_3H_6$) values of the M1-containing catalysts are much higher than those of the pure M2 phase at the same reaction temperature (e.g., 653 K). In particular, k" ($C_3H_6$) of the pure M1 phase is approximately three time higher than that of the pure M2 phase at 653 K, while their selectivity to ACN was similar This result clearly suggested that the M2 phase would yield less ACN than the M1 phase due to its lower activity in propylene ammoxidation (Figures 7 and 8). Therefore, these findings indicated the absence of synergy effect for the MoVSbNbO M1/M2 phases in the ammoxidation of propane to ACN because the MoVSbNbO M1 phase is significantly more efficient than the MoVSbNbO M2 phase in terms of converting the propylene intermediate to ACN.

*3.3.7 MoVSbTaO M1, M1/M2, and as-synthesized catalysts in propane ammoxidation*

The catalytic behaviors of the MoVSbTaO M1, M1/M2, and as-synthesized catalysts were investigated for the first time in propane ammoxidation to ACN. The selectivities of the MoVSbTaO M1, M1/M2, and as-synthesized catalysts during propane ammoxidation to ACN were plotted against propane conversion and shown in Figure 7 of Supporting Information. All MoVSbTaO catalysts showed similar maximum ACN selectivities suggesting the absence of the synergy effect for the MoVSbTaO system in propane ammoxidation.

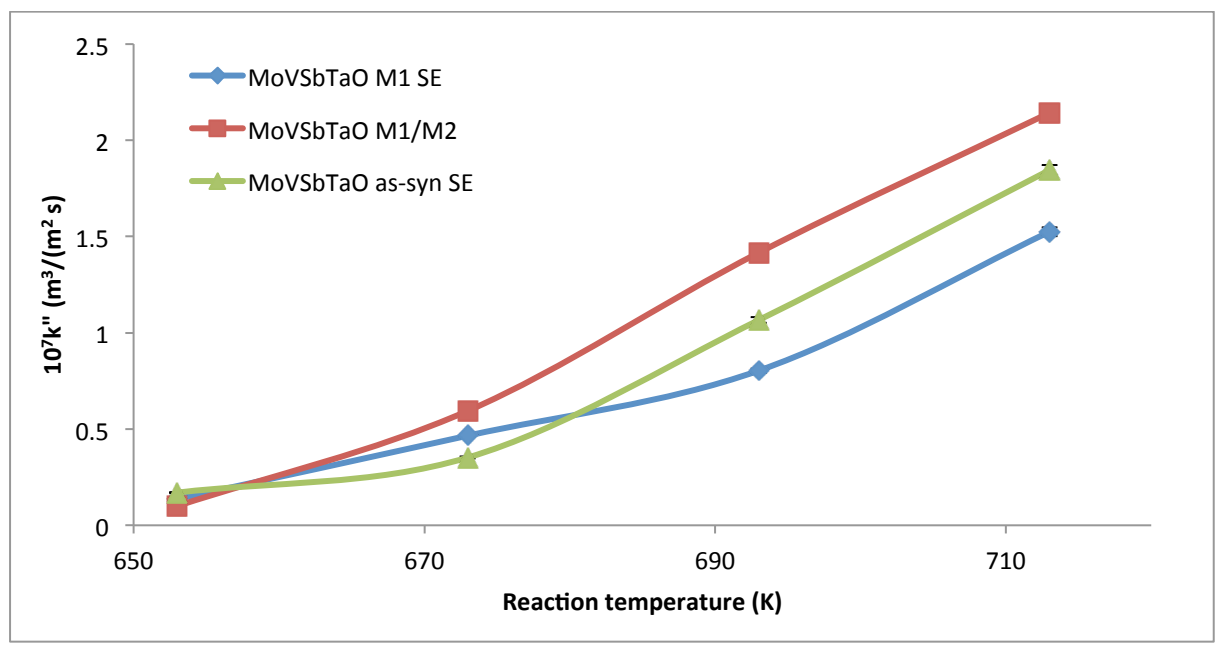

**Figure 16**. Reaction rate constants of propane consumption, k"($C_3H_8$), vs. reaction temperature for the MoVSbTaO M1, M1/M2, and as-synthesized catalysts in propane ammoxidation; Reaction conditions: $C_3H_8$:$NH_3$:$O_2$:He=5.7:8.6:17.1:68.6; total flow rate, 26.3 mL•min$^{-1}$; 0.2 g catalyst.

The k"($C_3H_8$) values of the MoVSbTaO M1, M1/M2, and as-synthesized catalysts are plotted against the reaction temperature in Figure 16 to provide further insights into catalytic behavior of newly reported MoVSbTaO system in propane ammoxidation. The grinding effect based on scenario 3 is likely to be responsible for the observed high k"($C_3H_8$) values of the

M1/M2 catalyst for propane ammoxidation shown in Figure 16. The low activity of the pure M1 phase as compared to the M1/M2 and as-synthesized catalysts (Figure 16) can be explained by removal of some surface active species from the M1 surface by the $H_2O_2$ treatment according to scenario 2.

*5.3.3.8 MoVSbTaO M1, M2, M1/M2, and as-synthesized catalysts in propylene ammoxidation*

The selectivities to ACN in propylene ammoxidation for the MoVSbTaO M1, M2, M1/M2, and as-synthesized catalysts were plotted against propylene conversion in Figure 17. It was found that the catalytic behavior of the MoVSbTaO M2 phase in propylene ammoxidation was very different from other M2 phases (Figure 8), and the M2 phase showed poor selectivity to ACN as compared to the M1 phase in propylene ammoxidation (Figure 17).

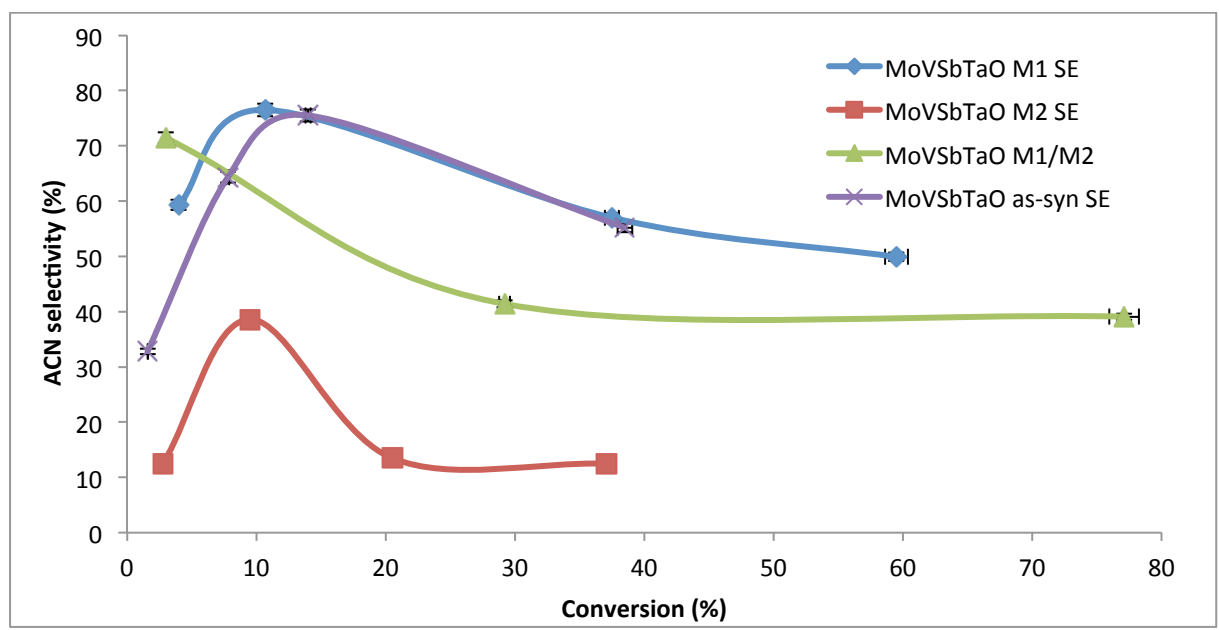

**Figure 17.** Selectivity to ACN as a function of propylene conversion over the MoVSbTaO M1, M2, M1/M2, and as-synthesized catalysts during propylene ammoxidation; Reaction conditions: $C_3H_6$:$NH_3$:$O_2$:He=5.7:8.6:17.1:68.6 (%); total flow rate, 26.3 mL•min$^{-1}$; 0.2 g catalyst; reaction temperature: 613 – 673 K.

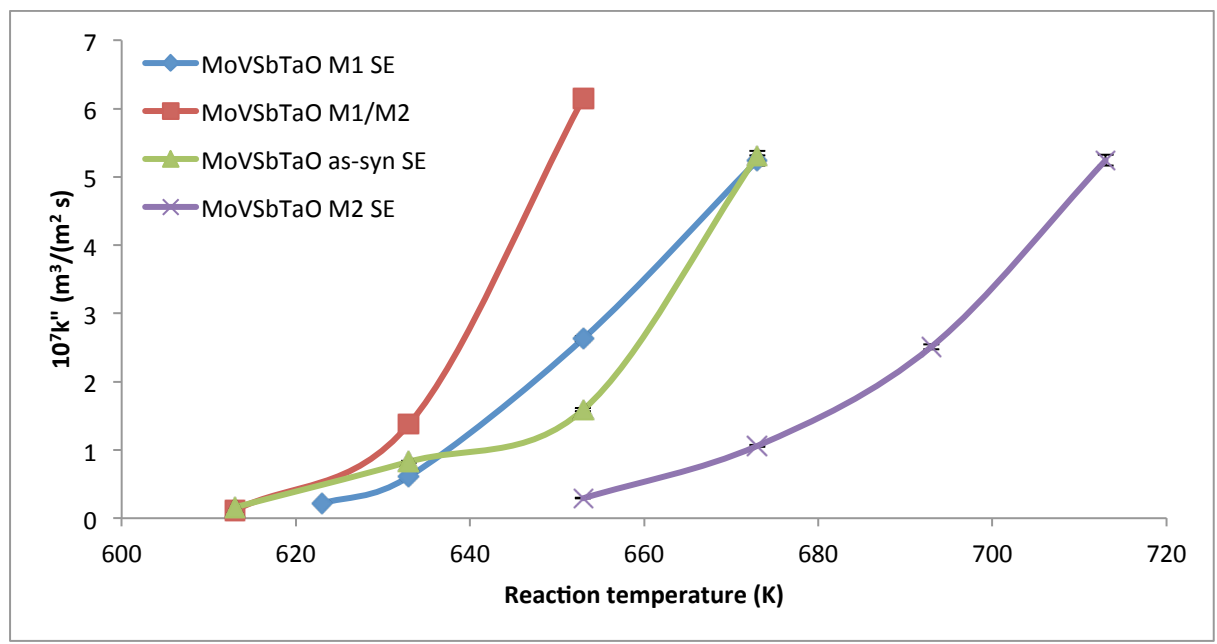

**Figure 18**. Reaction rate constants of propylene consumption, k" ($C_3H_6$), vs. reaction temperature for the MoVSbTaO M1, M2, M1/M2, and as-synthesized catalysts in propylene ammoxidation; Reaction conditions: $C_3H_6$:$NH_3$:$O_2$:He=5.7:8.6:17.1:68.6; total flow rate, 26.3 mL•min$^{-1}$; 0.2 g catalyst.

The k"($C_3H_6$) values of the MoVSbTaO M1, M1/M2, and as-synthesized catalysts are plotted against the reaction temperature in Figure 18, which shows that the MoVSbTaO M2 phase is significantly less active than the M1-containing catalysts at the same reaction temperature (e.g., 653 K). Therefore, the low ACN selectivity (Figure 17) and activity (Figure 18) of the M2 phase as compared to the M1 phase in propylene ammoxidation clearly suggested the lack of the M1/M2 synergy effect for the MoVSbTaO system in propane ammoxidation.

*3.3.9 MoVSbO M1, M1/M2, and as-synthesized catalysts in propane ammoxidation*

As mentioned in the introduction, the MoVSbO M2 phase showed poor selectivity toward acrylic acid contrary to that found for the MoVTeNbO system [27], suggesting the lack of synergy effect for the MoVSbO system in propane oxidation. In addition, the MoVSbO M1 and M2 phases have not been investigated for propane ammoxidation. Therefore, the MoVSbO

M1, M1/M2, and as-synthesized catalysts were probed for the synergy effect in propane ammoxidation. The selectivities to ACN for the MoVSbO M1, M1/M2, and as-synthesized catalysts are plotted as a function of reaction temperature in Figure 8 of Supporting Information. The M1/M2 and as-synthesized catalysts (Figure 8 of Supporting Information) showed much higher selectivity to ACN than the M1 phase.

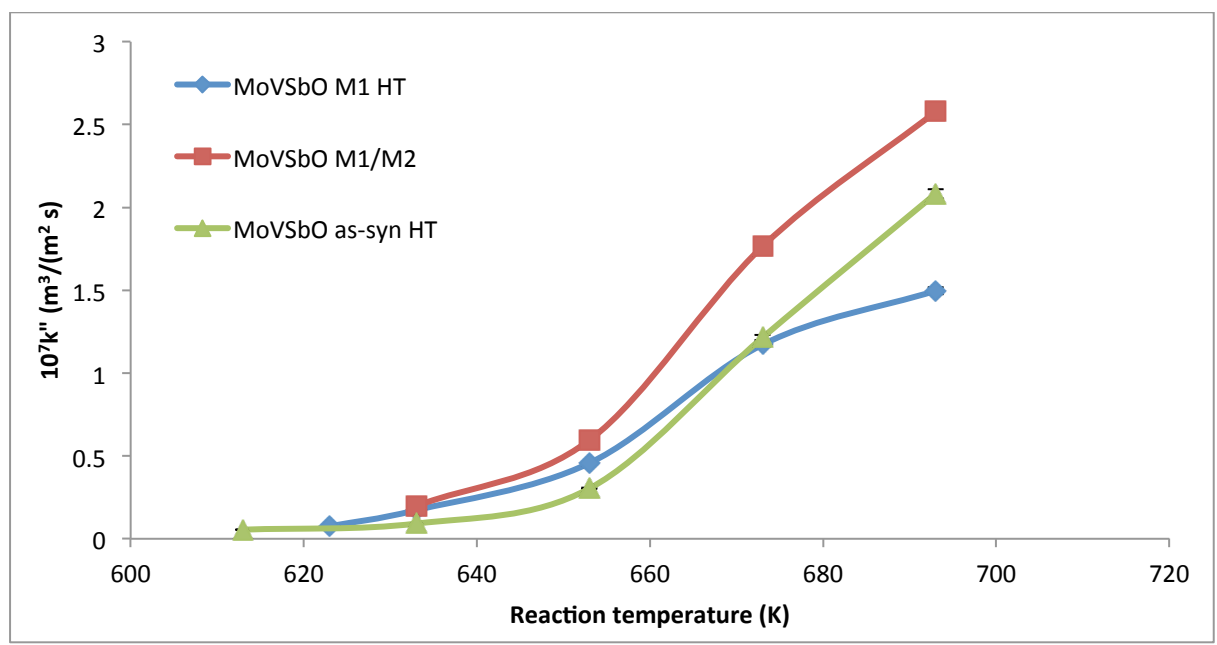

**Figure 19**. Reaction rate constants of propane consumption, k"($C_3H_8$), vs. reaction temperature for the MoVSbO M1, M1/M2, and as-synthesized catalysts in propane ammoxidation; Reaction conditions: $C_3H_8$:$NH_3$:$O_2$:He=5.7:8.6:17.1:68.6; total flow rate, 26.3 mL•min$^{-1}$; 0.2 g catalyst.

The k"($C_3H_8$) values of the MoVSbO M1, M1/M2, and as-synthesized catalysts are plotted as a function of reaction temperature in Figure 19. The MoVSbO M2 phase was tested and found to be inactive in propane ammoxidation similar to other M2 phases, which is consistent with the results of a previous study [27]. Figure 19 further showed that the M1/M2 catalyst containing 50 wt. % of inactive M2 phase displayed higher k"($C_3H_8$) values than the pure M1 phase. This result may be understood in terms of three possible scenarios described in section 3.3.1. Similar to that observed for other M1/M phase systems, the first scenario was excluded because the MoVSbO M2 phase is inactive towards propane. Therefore, the high k"($C_3H_8$) values of the M1/M2 catalyst in Figure 19 may be explained by the M1 phase grinding [33] according to scenario 3. This grinding effect may also account for the enhanced ACN selectivity for the M1/M2 catalyst in propane ammoxidation (Figure 8 of Supporting Information). The removal of some active species from the M1 surface due to the $H_2O_2$ treatment may explain the observed low k"($C_3H_8$) values of the pure M1 phase thus obtained.

*3.2.10 MoVSbO M1, M2, M1/M2, and as-synthesized catalysts in propylene ammoxidation*

Kinetic studies of the MoVSbO M1, M1/M2, and as-synthesized catalysts were performed in order to further understand higher selectivity to ACN for the M1/M2 and as-synthesized catalysts as compared to the pure M1 phase (Figure 8 of Supporting Information). The k"($C_3H_6$) values of the MoVSbO M1, M1/M2, and as-synthesized catalysts were plotted against the reaction temperature in Figure 20, which showed that the k"($C_3H_6$) of the M1/M2 catalyst is significantly higher than that of the pure M1 phase. The selectively exposed fresh surface *ab* planes resulted from the grinding of the M1 phase may be responsible for high k"($C_3H_6$) values of the M1/M2 catalyst as compared to the pure M1 phase in propylene ammoxidation because the MoVSbO M2 phase was found to be inactive for propylene ammoxidation in this study. Moreover, the relatively low k"($C_3H_6$) values of the pure M1 phase (Figure 20) may be explained by scenario 2 (detrimental impact of the $H_2O_2$ treatment). Therefore, these findings suggested that the observed greater ACN selectivity for the MoVSbO M1/M2 catalyst in propane ammoxidation (Figure 8 of Supporting Information) is not explained by the synergy effect but is due to mechanical grinding of the M1 phase (scenario 3).

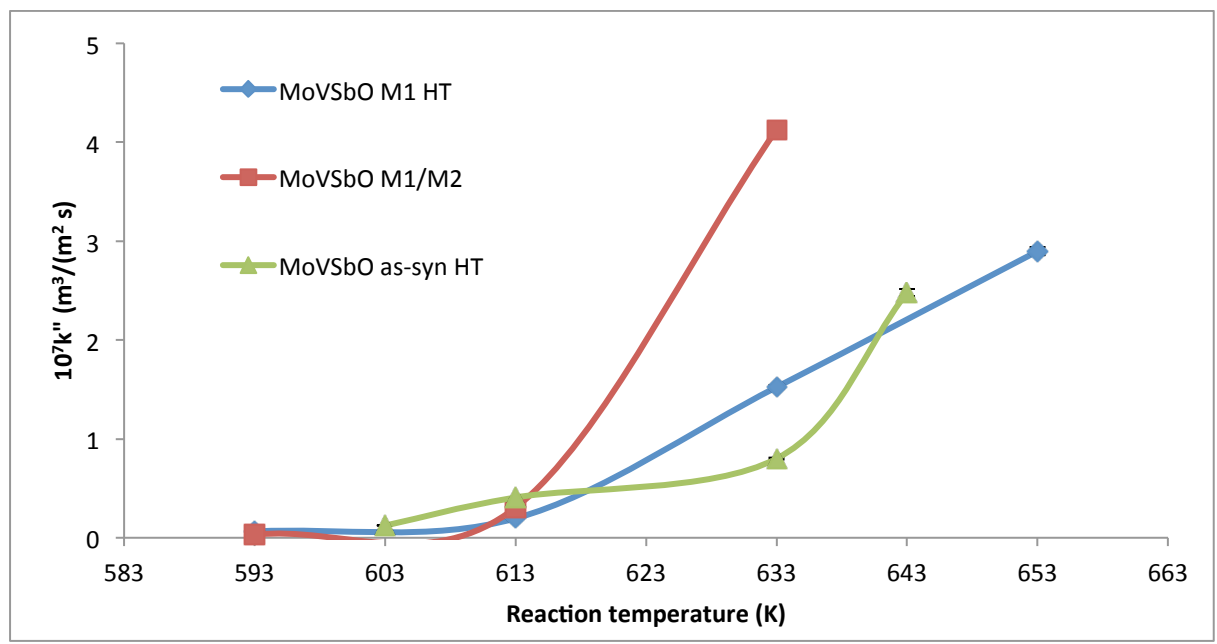

**Figure 20**. Reaction rate constants of propylene consumption, k" ($C_3H_6$), vs. reaction temperature for the MoVSbO M1, M2, M1/M2, and as-synthesized catalysts in propylene ammoxidation; Reaction conditions: $C_3H_6$:$NH_3$:$O_2$:He=5.7:8.6:17.1:68.6; total flow rate, 26.3 mL•min$^{-1}$; 0.2 g catalyst.

*3.4 Effects of $H_2O_2$ treatment and grinding on catalytic behavior of M1 phase*

The results of our kinetic studies indicated the lack of the synergy effect for all MoV(Te,Sb)(Nb,Ta)O M1/M2 phase variants, and suggested instead that the $H_2O_2$ treatment and mechanical grinding may be responsible for the observed reactivity trends in propane ammoxidation. The decrease of k"($C_3H_8$) of all pure MoV(Te,Sb)(Nb,Ta)O M1 phase catalysts is consistent with $H_2O_2$ treatment partially removing surface components from the M1 phase that are important for its activity in propane ammoxidation. For example, it may be further proposed that the $H_2O_2$ treatment may selectively remove some surface $TeO_x$ species that are known to be important for the catalytic activity and selectivity of the M1 phase towards propane (amm)oxidation given the labile nature of the TeOx species for the MoVTe(Nb,Ta)O systems. It was previously demonstrated that the catalytic activity and selectivity of the two-component

MoVO M1 phase was significantly enhanced when it was promoted with a submonolayer $TeO_x$ species introduced by the incipient wetness impregnation [45]. Recent theoretical studies from our group further indicated that the surface $TeO_x$ species may participate in both propane and propylene activation through α-hydrogen abstraction during propane (amm)oxidation [46]. Therefore, the partial Te loss from the surface of the MoVTe(Nb,Ta)O M1 phase may explain decreased activity and selectivity of the $H_2O_2$-treated MoVTe(Nb,Ta)O M1 phase in propane ammoxidation.

On the other hand, the recovery of catalytic activity after mechanical grinding of physical M1/M2 phase mixtures coincides with the generation of fresh surface *ab* planes of the M1 phase proposed to contain active and selective sites for propane ammoxidation [33]. Ueda et al. [47] found that grinding of Mo-V-M-O (M=Al, Ga, Bi, Sb, and Te) M1 phase catalysts, increased the conversion of propane and selectivity to acrylic acid in propane oxidation. Ohihara et al. [48] proposed that the grinding of catalysts is the most effective determinant for increased activity and selectivity to acrylic acid in propane oxidation. They concluded that the *ab* planes of the M1 phase are responsible for selective oxidation of propane to acrylic acid. In an earlier study from our group [43], the MoVTeNbO M1 phase was first coated by atomic layer deposition (ALD) method with alumina, which made it inactive in propane ammoxidation. This alumina-coated M1 phase was ground to expose fresh *ab* planes of the M1 phase which restored the catalytic activity, suggesting the *ab* planes of the M1 phase may be responsible for its high activity and selectivity in propane ammoxidation [43]. In another study from our group [33], crushing MoVTeTaO M1 phase significantly improved its catalytic reactivity in propane ammoxidation to ACN.

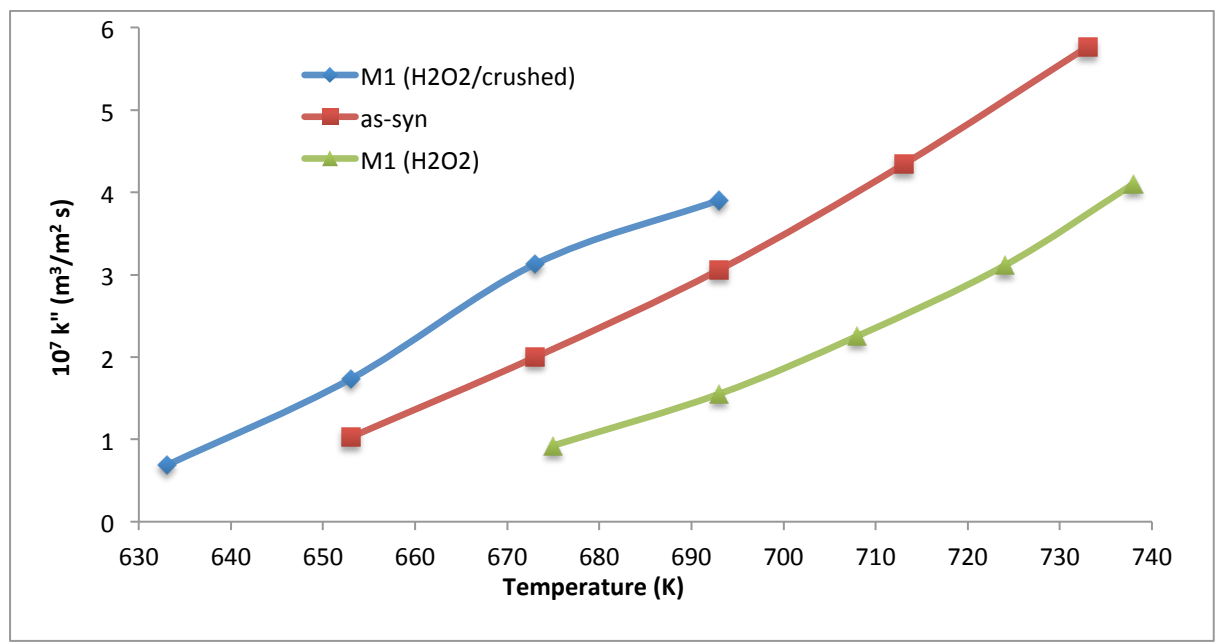

**Figure 21.** Reaction rate constants of propane consumption, k"($C_3H_8$), vs. reaction temperature of the as-synthesized MoVTeNbO [as-syn], the M1 phase after the $H_2O_2$ treatment [M1($H_2O_2$)], and M1 ground after the $H_2O_2$ treatment [M1($H_2O_2$/ground)] catalysts in propane ammoxidation; Reaction conditions: $C_3H_8$:$NH_3$:$O_2$:He=6:7:17:70; total flow rate, 20 mL•min$^{-1}$; 0.2 – 0.4 g catalyst.

The effect of the $H_2O_2$ treatment and subsequent grinding of the M1 phase was recently further investigated by our group for the MoVTeNbO system [49]. The reaction rate constants, k"($C_3H_8$), of propane consumption for the as-synthesized MoVTeNbO catalyst containing a mixture of the M1 and M2 phases, pure MoVTeNbO M1 phase after the $H_2O_2$ treatment, and MoVTeNbO M1 phase *ground* after the $H_2O_2$ treatment are shown in Figure 21. The detailed information about these catalysts is summarized in Table 1 of Supporting Information. The activity of the M1 phase decreased after the $H_2O_2$ treatment as compared to the as-synthesized catalyst. On the other hand, grinding of the M1 phase after the $H_2O_2$ treatment not only restored the catalytic activity, but further enhanced it as compared to the as-synthesized catalyst. The activation energies of propane consumption estimated from the Arrhenius plots of k"($C_3H_8$) of these three catalysts were in the 103-108 kJ/mol range indicating the similar nature of propane-activating sites in these catalysts.

**Table 3**. Elemental compositions and BET surface areas of the MoVTe(Nb,Ta)O M1 phase catalysts and elemental compositions of used $H_2O_2$ treatment solutions.

| Samples | Preparative[a] Mo/V/Te/Ta | ICP-MS Mo/V/Te/Ta | S.A. ($m^2/g$)[b] Before $H_2O_2$ treatment[c] | S.A. ($m^2/g$)[b] After $H_2O_2$ treatment[d] |
|---|---|---|---|---|
| M1 Nb HT $H_2O_2$ | | 1/0.28/0.13/0.07[e] | | |
| M1 HT (0.09) | 1.00/0.31/0.22/0.09 | 1/0.32/0.11/0.22 | 22.4 | 37.2 |
| M1 HT (0.09) $H_2O_2$ | | 1/0.28/0.14/0.00 | | |
| M1 HT (0.12) $H_2O_2$ | | 1/0.26/0.09/0.04[e] | | |
| M1 HT (0.15) | 1.00/0.31/0.22/0.15 | 1/0.38/0.13/0.23 | 4.9 | 12.3 |
| M1 HT (0.15) $H_2O_2$ | | 1/0.25/0.20/0.20 | | |
| M1 MW | 1.00/0.31/0.22/0.12 | 1/0.25/0.14/0.35 | 3.1 | 17.7 |
| M1 MW $H_2O_2$ | | 1/0.26/0.29/0.14 | | |

[a] Synthesis compositions; [b] S.A. = BET surface areas; [c] for as-synthesized catalysts; [d] for pure M1 phases; [e] corresponding M1 phase composition shown in Table 2; MW (Microwave-assisted hydrothermal synthesis); HT (Hydrothermal synthesis)

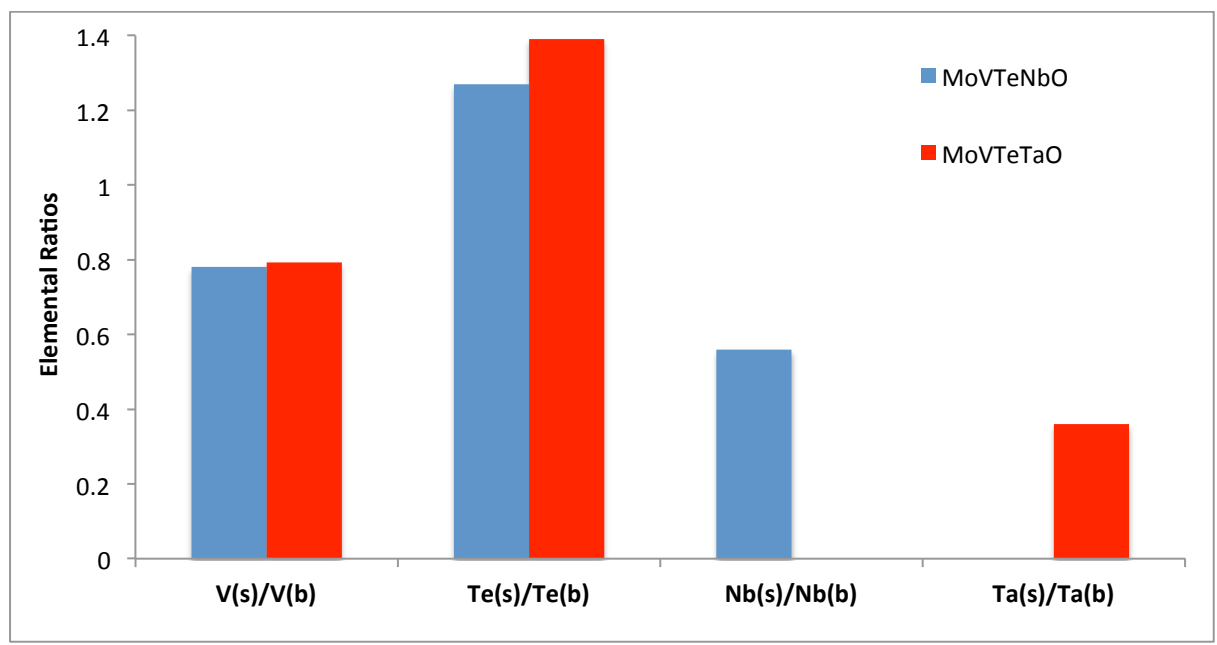

**Figure 22.** Elemental ratios in $H_2O_2$ solutions (s) referenced to elemental ratios in the $H_2O_2$-treated M1 phases (b) (Tables 2 and 3).

The ground MoVTeNbO M1 phase catalyst was further treated by aqueous 30% $H_2O_2$ as described in the experimental section in order to elucidate the nature of metal oxide species removed from the M1 phase by $H_2O_2$. The pure MoVTeTaO M1 catalysts (HT 0.09, HT 0.12, HT 0.15 and MW) were also ground and treated by $H_2O_2$. The details of preparation of these catalysts were described in our earlier study [28]. The $H_2O_2$ solutions after this treatment were separated from the solid catalyst by centrifugation and analyzed by ICP-MS (Table 3). These metal concentrations were converted to metal ratios by normalizing them to their Mo concentrations, which were in turn divided by their content in the bulk M1 phases in order to determine the nature of metal species preferentially leached by $H_2O_2$. These normalized ratios are shown in Figure 22 indicating that V and especially Ta and Nb were relatively depleted in the $H_2O_2$ solution as compared to Te. The average elemental M(s)/M(b) ratio for the four MoVTeTaO M1 phases were 1.39 for Te, 0.79 for V, and only 0.36 for Ta. These conclusions are further supported by the ICP-MS analysis of three other MoVTeTaO M1 phases shown in Table 3 that were treated similarly. The average elemental M(s)/M(b) ratio for the four MoVTeTaO M1 phases were 1.39 for Te, 0.79 for V, and only 0.36 for Ta. These findings confirmed the hypothesis that the $H_2O_2$ treatment indeed removes metal oxides species from the M1 phase and does so preferentially towards the $TeO_x$, which is detrimental for its catalytic activity and selectivity in propane (amm)oxidation.

## 4. Conclusions

In this study, we systematically explored the catalytic activity and selectivity of the MoV(Te,Sb)(Nb,Ta)O M1 and M2 phase catalysts prepared by the slurry evaporation (SE) and hydrothermal synthesis (HT) methods in propane ammoxidation. For the very first time, the MoVSbTaO M1 and M2 phases were synthesized and characterized. It was found that the newly synthesized MoVSbTaO M1 and M2 phase are also active and selective in propane (M1 phase only) and propylene (M1 and M2 phase) ammoxidation. This study confirmed that the MoVTeNbO M1 phase is the best catalyst for propane ammoxidation because of its high activity and selectivity to ACN. It was also found that the M2 phases of all compositions investigated here were active in propylene ammoxidation except the MoVSbO M2 phase, but displayed different selectivities to ACN depending on chemical composition.

Most importantly, the kinetic study of the MoV(Te,Sb)(Nb,Ta)O M1 and M2 phases in propylene ammoxidation revealed for the very first time that the M2 phases are significantly less active than the corresponding M1 phases in propylene ammoxidation. The findings of this study do not support the existence of the synergy effect for any MoV(Te,Sb)(Nb,Ta)O M1/M2 system. Instead, the observed trends of the MoV(Te,Sb)(Nb,Ta)O catalysts in propane ammoxidation were consistent with partial loss of some surface active species from the surface of the M1 phase during the $H_2O_2$ treatment and generation of fresh *ab* planes of the M1 phase via mechanical grinding of the $H_2O_2$ treated M1 phase. These findings provided further evidence that the M1 phase is the only phase required for the activity and selectivity of the MoV(Te,Sb)(Nb,Ta)O catalysts in propane ammoxidation to ACN.

## 5. Acknowledgments

This study was supported by the Chemical Sciences, Geosciences and Biosciences Division, Office of Basic Energy Sciences, U.S. Department of Energy, under Grant #DE-FG02-04ER15604. The author is grateful to Dr. Korovchenko (University of Cincinnati) for the catalytic characterization of MoVTeNbO catalysts shown in Figure 21.

# Supporting Information

**Table S1.** Elemental composition and BET surface areas of the MoVTeNbO M1 phase catalysts.

| Catalysts | Preparative [a] (Mo/V/Te/Nb) | Surface area ($m^2/g$) [b] |
|---|---|---|
| as-synthesized MoVTeNbO | 1/0.3/0.17/0.12 | 9.8 |
| M1 phase after the $H_2O_2$ treatment | | 10.1 |
| M1 ground after the $H_2O_2$ treatment | | 14 |

[a] Synthesis composition in the slurry; [b] measured by the BET method.

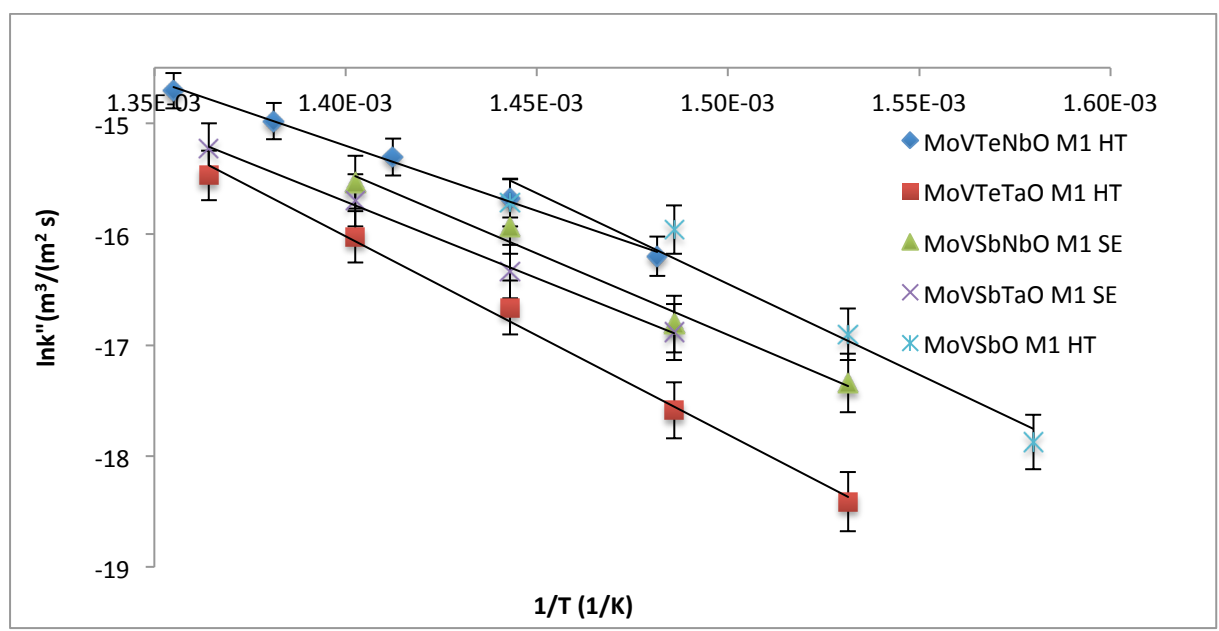

**Figure S1.** Arrhenius plots of k" of propane consumption over the MoV(Te,Sb)(Nb,Ta)O M1 phase catalysts in propane ammoxidation; Reaction conditions: $C_3H_8$:$NH_3$:$O_2$:He=5.7:8.6:17.1:68.6; total flow rate, 26.3 mL•$min^{-1}$; 0.2 g catalyst; reaction temperature: 623-733 K.

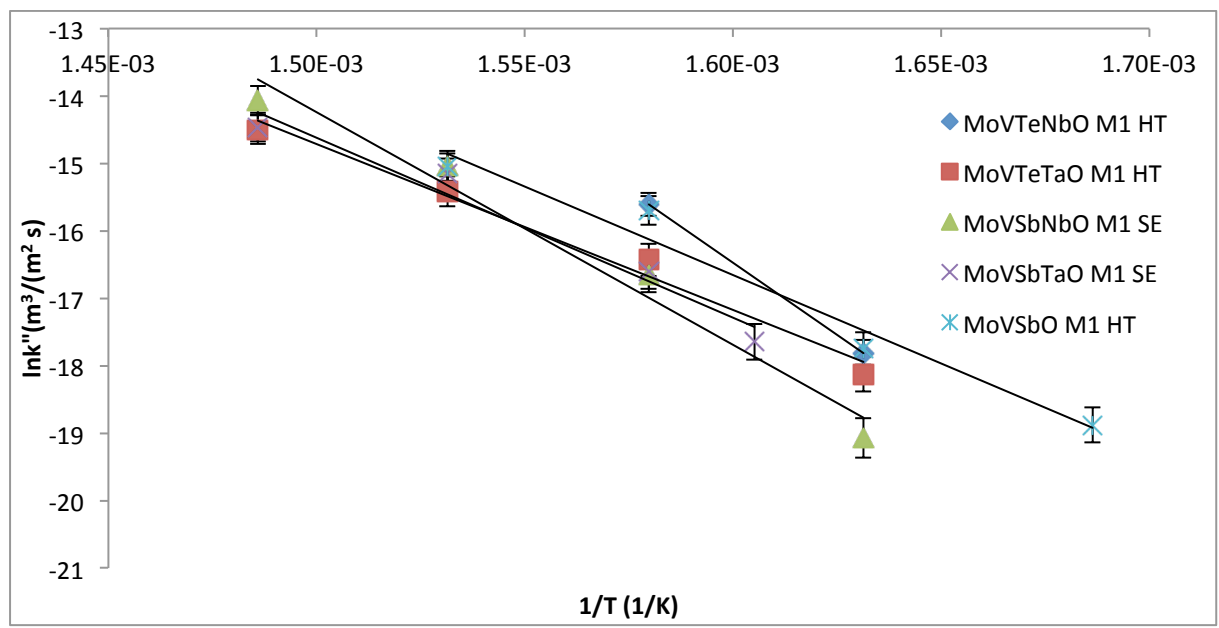

**Figure S2.** Arrhenius plots of k" of propylene consumption over the MoV(Te,Sb)(Nb,Ta)O M1 phase catalysts in propylene ammoxidation; Reaction conditions: $C_3H_8:NH_3:O_2:He=5.7:8.6:17.1:68.6$; total flow rate, 26.3 mL•min$^{-1}$; 0.2 g catalyst; reaction temperature: 623-733 K.

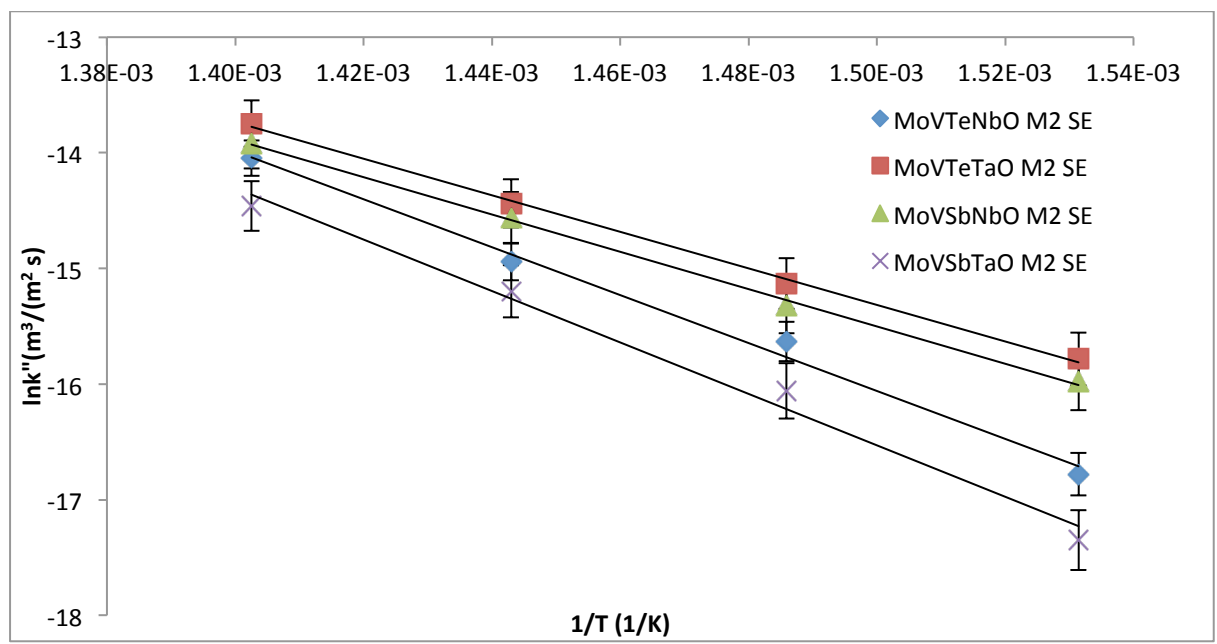

**Figure S3.** Arrhenius plots of k" of propylene consumption over the MoV(Te,Sb)(Nb,Ta)O M2 phase catalysts in propylene ammoxidation; Reaction conditions: $C_3H_8:NH_3:O_2:He=5.7:8.6:17.1:68.6$; total flow rate, 26.3 mL•min$^{-1}$; 0.2 g catalyst; reaction temperature: 623-733 K.

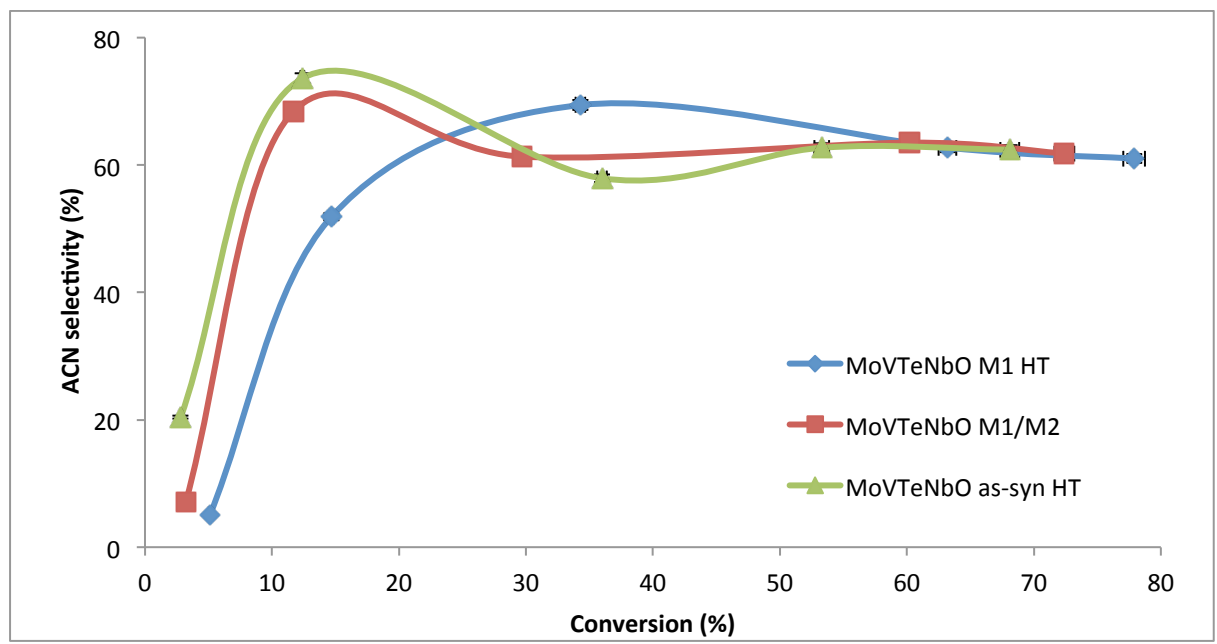

**Figure S4.** Selectivity to ACN as a function of propane conversion over the MoVTeNbO M1, M1/M2, and as-synthesized catalysts during propane ammoxidation; Reaction conditions: $C_3H_8:NH_3:O_2:He=5.7:8.6:17.1:68.6$ (%); total flow rate, 26.3 mL•min$^{-1}$; 0.2 g catalyst; reaction temperature: 633-713 K.

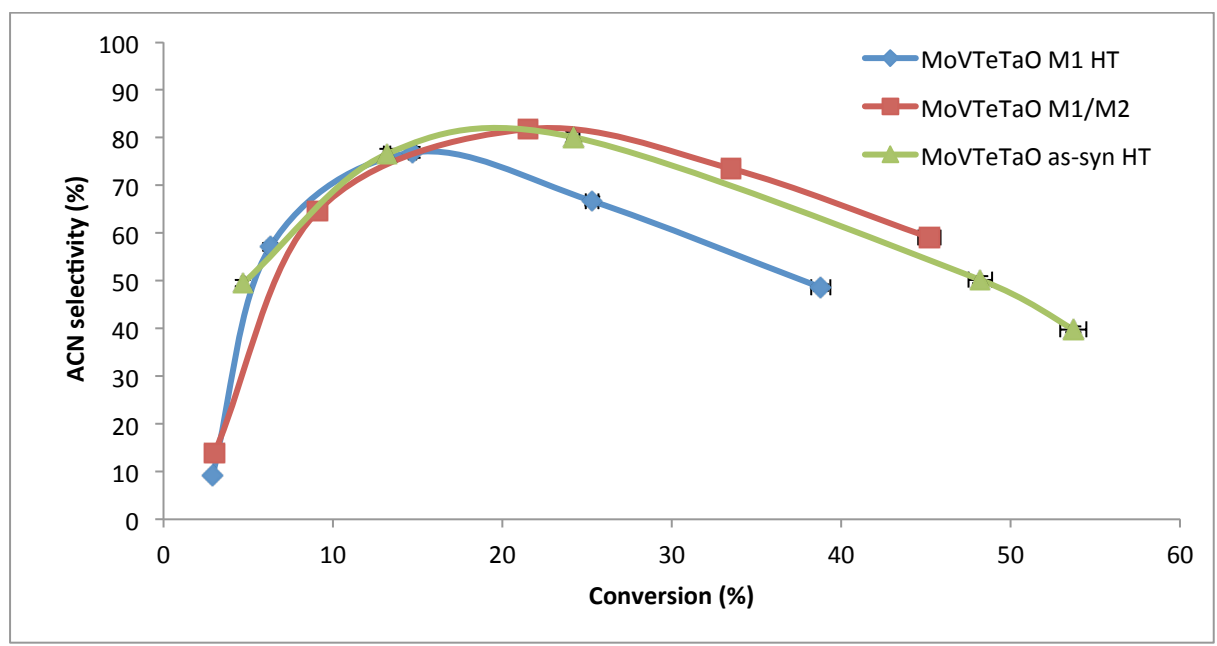

**Figure S5.** Selectivity to ACN as a function of propane conversion over the MoVTeTaO M1, M1/M2, and as-synthesized catalysts during propane ammoxidation; Reaction conditions: $C_3H_8:NH_3:O_2:He=5.7:8.6:17.1:68.6$ (%); total flow rate, 26.3 mL•min$^{-1}$; 0.2 g catalyst; reaction temperature: 653-733 K.

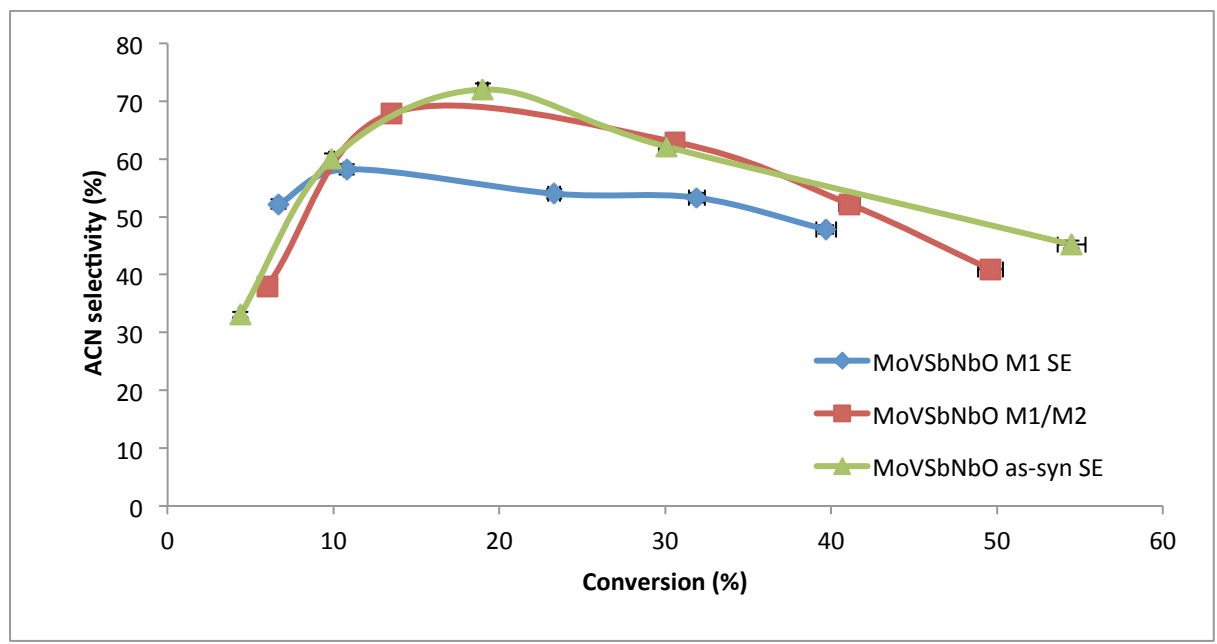

**Figure S6.** Selectivity to ACN as a function of propane conversion over the MoVSbNbO M1, M1/M2, and as-synthesized catalysts during propane ammoxidation; Reaction conditions: $C_3H_8:NH_3:O_2:He=5.7:8.6:17.1:68.6$ (%); total flow rate, 26.3 mL•min$^{-1}$; 0.2 g catalyst; reaction temperature: 653-733 K.

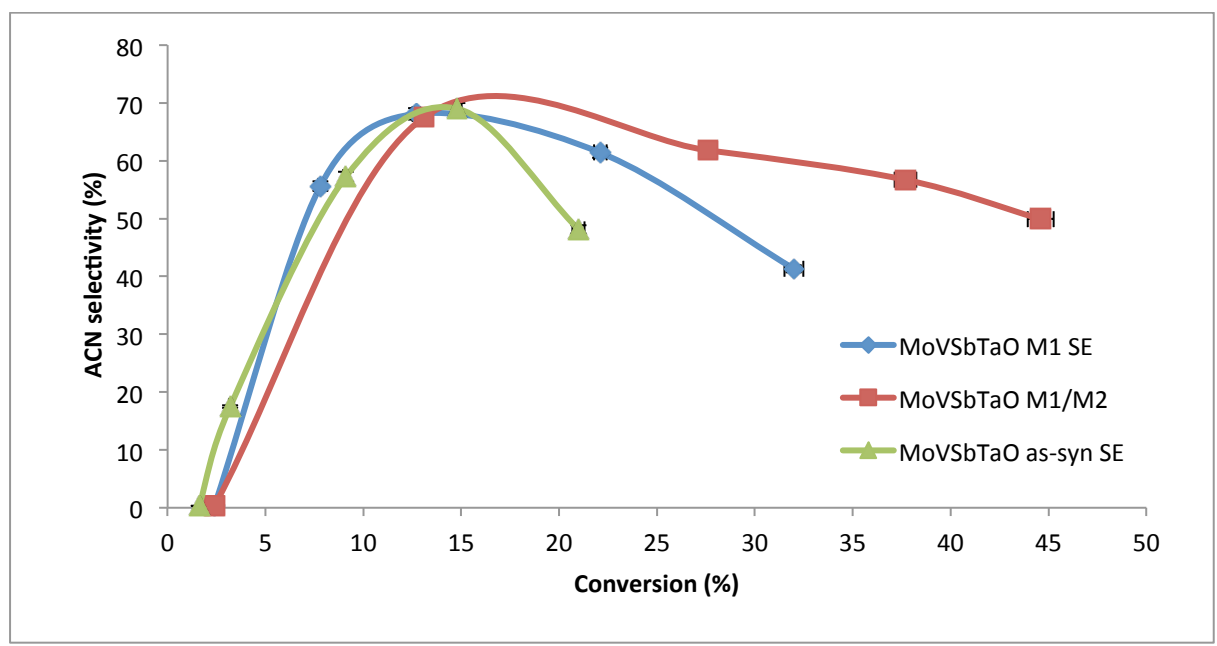

**Figure S7.** Selectivity to ACN as a function of propane conversion over the MoVSbTaO M1, M1/M2, and as-synthesized catalysts during propane ammoxidation; Reaction conditions: $C_3H_8:NH_3:O_2:He=5.7:8.6:17.1:68.6$ (%); total flow rate, 26.3 mL•min$^{-1}$; 0.2 g catalyst; reaction temperature: 653-733 K.

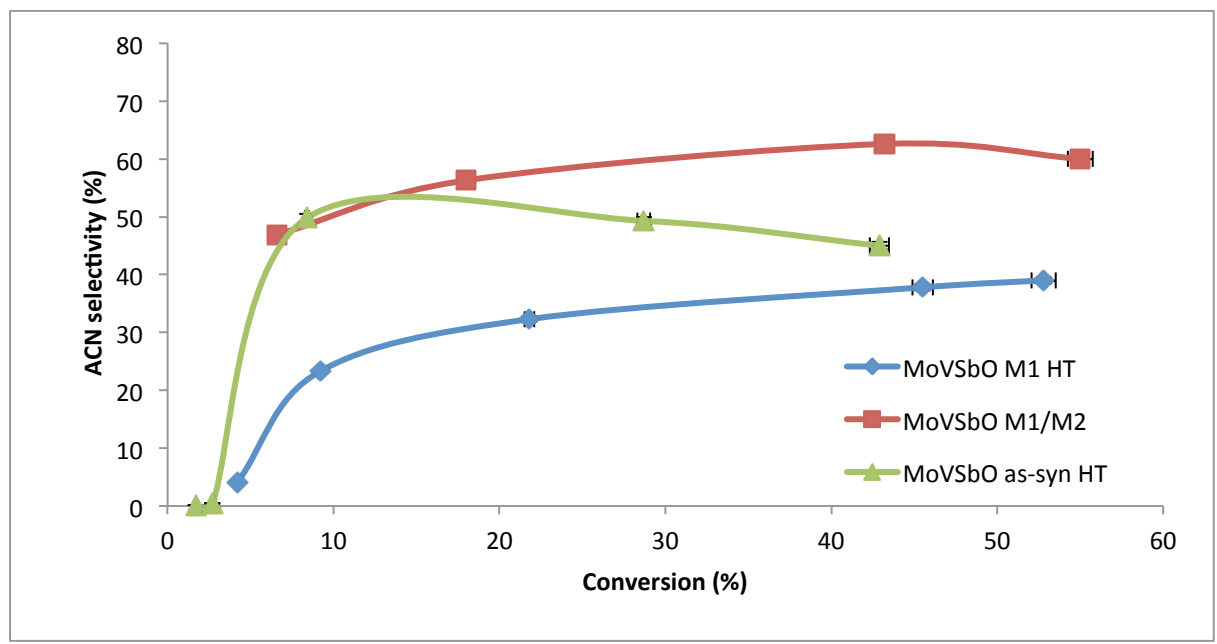

**Figure S8.** Selectivity to ACN as a function of propane conversion over the MoVSbO M1, M1/M2, and as-synthesized catalysts during propane ammoxidation; Reaction conditions: $C_3H_8:NH_3:O_2:He=5.7:8.6:17.1:68.6$ (%); total flow rate, 26.3 mL•min$^{-1}$; 0.2 g catalyst; reaction temperature: 623-693 K.